\documentclass[12pt,a4paper]{article}

\usepackage{amssymb}
\usepackage{amsmath}
\usepackage{amsthm}
\usepackage[usenames]{color}

\newtheorem{theorem}{Theorem}

\newtheorem{definition}{Definition}

\newtheorem{lemma}{Lemma}

\newtheorem{proposition}{Proposition}
\newtheorem{remark}{Remark}

\usepackage{pgfplots}
\pgfplotsset{compat=1.15}
\usepackage{mathrsfs}
\usetikzlibrary{arrows}

\begin{document}

\begin{center}
\quad \bigskip
\quad \bigskip

{\Large Obvious Strategy-proofness with Respect to a Partition}\nolinebreak 
\renewcommand{\thefootnote}{
\fnsymbol{footnote}}\footnote{
We acknowledge financial support from the grants PID2020-116771GB-I00 and PID2023-147183NB-I00 founded by MCIN/AEI/10.13039/501100011033.
Arribillaga and Neme acknowledge financial support from the UNSL (grants 032016 and 032016), the CONICET (grant PIP112-200801-00655), and the Agencia I+D+i (grant PICT2017-2355).
Mass\'{o} acknowledges financial support from the Spanish Agencia Estatal de Investigaci\'{o}n (AEI), through the Severo Ochoa Programme for Centers of Excellence in R\&D (Barcelona School of Economics CEX2019-000915-S) and the Generalitat de Catalunya, through the grant 2021SGR-00194.\smallskip}\bigskip

\quad \medskip

\textsc{R. Pablo Arribillaga}\footnote{Instituto de Matem\'{a}tica Aplicada San Luis, Universidad Nacional de San Luis and CONICET. 
Ej\'{e}rcito de los Andes 950. 
5700, San Luis, Argentina (e-mails: rarribi@unsl.edu.ar and aneme@unsl.edu.ar).\smallskip}\qquad \textsc{Jordi Mass\'{o}}\footnote{Corresponding author. 
Universitat Aut\`{o}noma de Barcelona and Barcelona School of Economics. 
Departament d'Economia i Hist\`{o}ria Econ\`{o}mica.
Edifici B, Campus UAB. 
08193, Cerdanyola del Vallès (Barcelona), Spain (e-mail: jordi.masso@uab.es).\smallskip}\qquad \textsc{Alejandro Neme}{\small \thinspace }$^{\dagger }\bigskip $

July 26, 2024\bigskip
\end{center}

\setcounter{footnote}{0}

\begin{quote}
\noindent {\small {\ \underline{Abstract}: We define and study obvious strategy-proofness with respect to a partition of the set of agents. 
It encompasses strategy-proofness as a special case when the partition is the coarsest one and obvious strategy-proofness when the partition is the finest. 
For any partition, it falls between these two extremes. 
We establish two general properties of this new notion and apply it to the simple anonymous voting problem with two alternatives and strict preferences.}

\textsc{Keywords:} Obvious strategy-proofness; Extended majority voting.\\
\textsc{JEL Classification Number:} D71. }
\end{quote}

\section{Introduction}

We propose and investigate a novel implementation concept termed \textit{obvious strategy-proofness with respect to a partition}.
This concept is stronger than strategy-proofness and weaker than obvious strategy-proofness as defined by Li (2017). 
It aligns with strategy-proofness when considering the coarsest partition and with obvious strategy-proofness for the finest partition.

A social choice function, which maps preference profiles into the set of alternatives, is strategy-proof if truth-telling is always an optimal decision for all agents.
Li (2017) argues that strategy-proofness requires that agents are able to perform complex contingent reasoning: 
Each agent must be capable of recognizing, for every conceivable profile of preferences declared by other agents (the contingencies it confronts when determining its own declared preference), that truth-telling stands among the optimal choices.

To ease the cognitive load on agents' reasoning, Li (2017) proposes substituting the hypothetical contingencies of the simultaneous mechanism (\textit{i.e.}, a normal game form) with verifiable facts in a sequential mechanism (\textit{i.e.}, an extensive game form). 
These facts can be observed by the agent at any point when it needs to make a decision throughout the extensive game form.
Furthermore, to evaluate the consequence of truth-telling against any alternative choice, a behavioral hypothesis is employed regarding the future actions of all subsequent players: it adopts a pessimistic stance when assessing truth-telling (anticipating the worst possible future outcome) and an optimistic outlook when evaluating deviations (expecting the best possible future outcome). 
If the worst outcome linked to truth-telling is at least as good as the best outcome linked to any deviation, then truth-telling emerges as an evidently optimal choice, or in other words, an obviously dominant strategy.

Numerous papers have delved into the study of obvious strategy-proofness.
For a general setting, see for instance Bade and Gonczarowski (2017), Mackenzie (2020), Mackenzie and Zhou (2022), and Pycia and Troyan (2023). 
For more specific contexts addressing particular facets of obvious strategy-proofness, see for instance Arribillaga, Massó and Neme (2020 and 2023), Ashlagi and Gonczarowski (2018), Ferraioli, Meier, Penna and Ventre (2019), Ferraioli and Ventre (2019), Golowich and Li (2022), Tamura (2024), Thomas (2021), and Troyan (2019).

For a given partition of the set of agents, our notion blends elements from both extremes, while preserving the sequential interpretation of the direct revelation mechanism.\footnote{The direct revelation mechanism is a normal game form that can also be described as an extensive game form with imperfect information.
In this game, agents play only once, without knowledge of the choices made by other agents, by choosing a preference from the set of their preferences.
The outcome associated with each terminal node is the alternative that the social choice function would select at the profile of preferences chosen along the path leading to that node.\smallskip} 
Given a partition of the set of agents, each agent, when making a choice, assumes that the strategies of other agents within the same subset of the partition are fixed and taken as given\textemdash a hypothetical contingency. 
Meanwhile, the agent uses the two most extreme behavioral hypotheses to evaluate the future choices of agents outside its subset who must play thereafter.
It can be considered easier to perform contingent reasoning about the choices of agents belonging to the same subset of the partition than about those in other subsets.
For instance, agents within the same subset may engage in pre-play communication and develop a shared hypothesis about the choices that subset members will make throughout the game. 
Therefore, it is reasonable to regard the behavior of agents within the same subset of the partition as both hypothetical and given when evaluating one's own choice.
In contrast, information about agents outside one's own subset may be limited, and pre-play communication may not be feasible. Therefore, when comparing truth-telling with deviating at the moment of making the choice, the agent may not be able to determine the subsequent actions of agents outside its own subset.
Consequently, it may consider their choices as unfixed and instead rely on extreme assumptions about their potential consequences.

Our main results are as follows.
Firstly, Proposition \ref{Elevator} shows that if a social choice function can be implemented in obviously dominant strategies with respect to a partition, then it can also be implemented in obviously dominant strategies with respect to any coarser partition. 
Secondly, for any partition of the set of agents, Theorem \ref{SP-implements} identifies a broad and straightforward class of extensive game forms.
We show that if a social choice function can be implemented in dominant strategies by a game within this class, then it can also be implemented in obviously dominant strategies with respect to the partition by the same game.
Furthermore, in Remark \ref{RTM} we argue that to implement a social choice function using obviously dominant strategies with respect to a partition, it suffices to focus solely on extensive game forms belonging to a class that plays a significant role in the literature on obvious strategy-proofness: round table mechanisms (see Mackenzie (2020)).

The paper then applies the new implementation concept of obvious strategy-proofness with respect to a partition to the simplest social choice problem involving only two alternatives, denoted as $x$ and $y$, with agents' preferences being strict.
This simple setting admits a broad family of strategy-proof social choice functions, called \textit{extended majority voting rules}.
Each rule within this class can be described as a committee comprising a monotonic family of winning coalitions. 
These winning coalitions represent subsets of agents capable of ensuring the selection of alternative $x$ by voting for it, irrespective of the other agents' votes.
We aim to identify all extended majority voting rules that are obviously strategy-proof with respect to a partition.
Our analysis is restricted to the anonymous family. See Arribillaga, Massó and Neme (2024a) for the analysis of the general non-anonymous case.
We characterize two nested families of extended majority voting rules, each of which is obviously strategy-proof with respect to a partition. 
These families correspond to two anonymous subclasses associated with two distinct notions of anonymity. 
These notions differ in terms of the types of permutations of the set of agents for which the extended majority voting rules are required to remain invariant.
\textit{Anonymity relative to a partition} allows only permutations that map each subset of the partition into itself, ensuring that the partition remains unchanged by the permutation.
\textit{Strong anonymity} permits agents to be permuted in any manner, allowing a partitioned set of agents to be mapped into a potentially different partition.
Theorems \ref{anonymity} and \ref{SA} contain these two characterizations.
In Arribillaga, Massó and Neme (2024a), we address the general case without anonymity and identify a necessary and sufficient condition, called the \textit{Iterated Union Property} (IUP), that a committee must satisfy for the corresponding extended majority voting rule to be implementable by means of a simple extensive game form in obviously dominant strategies with respect to a partition.

The paper is organized as follows. 
Section 2 presents the basic notation, definitions and the description of extensive game forms necessary to define obvious strategy-proofness with respect to a partition, which is introduced and studied in Section 3. 
Section 4 applies this new notion involving two alternatives and strict preferences, characterizing two nested families of anonymous and obviously strategy-proof social choice functions with respect to a partition.
An Appendix contains the proof of a lemma omitted in the main text.

\section{Preliminaries}

\subsection{Basic notation and definitions}

We consider collective decision problems where a set of agents $N=\{1,\dots,n\}$ must select an alternative from a given set $A$. 
Each agent $i\in N$ has a (weak) preference $R_i$ over $A$, which is a complete and transitive binary relation on $A$. 
For a given preference $R_i$, we denote by $P_i$ its induced strict preference, and by $t(R_i)$ the most-preferred alternative according to $R_i$, if it exists. 
Specifically, for any distinct pair $x,y\in A$, $x\,P_i\,y$ if and only if $x\,R_i\,y$ and not $y\,R_i\,x$, and $t(R_i)\,P_i\,y$ for all $y\in A\setminus \{t(R_i)\}$. 
Let $\mathcal{R}$ and $\mathcal{P}$ denote the sets of all weak and strict preferences over $A$, respectively. 
A (preference) \emph{profile} is defined as an $n$-tuple $R=(R_1,\ldots ,R_n)\in \mathcal{R}^N$, representing an ordered list of $n$ preferences, one for each agent in the set $N$.
Given a profile $R$, an agent $i$, and a subset of agents $S$, we denote by $R_{-i}$ and $R_{-S}$ the sub-profiles in $\mathcal{R}^{N\setminus\{i\}}$ and $\mathcal{R}^{N\setminus S}$ obtained by removing $R_i$ and $R_{S}:=(R_j)_{j\in S}$ from $R$, respectively.
Therefore, $R$ can be represented as $(R_i,R_{-i})$ or as $(R_S,R_{-S})$.

A \emph{social choice function} $f:\mathcal{D}\rightarrow A$ on a Cartesian product domain of preference profiles $\mathcal{D}:=\mathcal{D}_1 \times \cdots \times \mathcal{D}_n\subseteq \mathcal{R}^N$ selects, for each profile $R\in \mathcal{D}$, an alternative $f(R)\in A$.
Given a social choice function $f:\mathcal{D}\rightarrow A$, we refer to an agent $i\in N$ as a \textit{dummy} agent in $f$ if, for all $R\in \mathcal{D}$ and all $R^\prime _i \in \mathcal{D}_i$,
\[
f(R_i,R_{-i})=f(R^\prime_i,R_{-i});
\]
namely, $f$ remains invariant with respect to agent $i$'s preference.

Let $f:\mathcal{D}\rightarrow A$ be a social choice function. 
We construct its associated normal game form $(N,\mathcal{D},f)$, where $N$ is the set of players, $\mathcal{D}$ is the Cartesian product set of strategy profiles and $f$ is the outcome function mapping strategy profiles into alternatives.
Then, $f$ is implementable in dominant strategies (or $f$ is SP-implementable) if the normal game form $(N,\mathcal{D},f)$ has the property that, for all $i\in N$ and $R\in \mathcal{D}$, $R_{i}$ is a weakly dominant strategy for $i$ in the game in normal form $(N,\mathcal{D},f,R)$, where each $i\in N$ uses $R_{i}$ to compare the consequences of pairs of strategy profiles.
The literature refers to $(N,\mathcal{D},f)$ as the direct revelation mechanism that SP-implements $f$.

Equivalently, a social choice function $f:\mathcal{D}\rightarrow A$ is \emph{strategy-proof} (SP) if, for all $i\in N$, $R\in \mathcal{D}$, and $R^{\prime }_i\in \mathcal{D}_i$, 
\begin{equation*}
f(R_i,R_{-i})\,R_i\,f(R^{\prime }_i,R_{-i}).
\end{equation*}

Strategy-proofness requires that agents are capable of engaging in contingent reasoning, which can be complex, even for relatively simple social choice functions. 
To accommodate agents who may have limited abilities in this regard, Li (2017) introduces the stronger incentive notion of obvious strategy-proofness (OSP) for general settings wherein agents' types (coinciding with agents' preferences in our setting) are considered private information.
A social choice function $f:\mathcal{D}\rightarrow A$ is \emph{obviously strategy-proof} (OSP) if it satisfies two conditions. 
First, there must exist an extensive game form $\Gamma $, played by the agents in $N$, with outcomes corresponding to the alternatives in $A$. 
Additionally, there must be a type-strategy profile $(\sigma_i^{R_i})_{R_i\in \mathcal{D}_i\,,\,i\in N}$, a behavioral strategy in $\Gamma$ for each agent and for each of its types (to be formally defined in Subsection 2.2), that induces the social choice function.
Namely, for every profile of types $R=(R_1,\dots, R_n)\in \mathcal{D}$, when each agent $i$ plays the strategy $\sigma^{R_i}_i$ that corresponds to its type $R_i$, the outcome of the game $x$ is the alternative that the social choice function would have chosen at this profile (\textit{i.e.}, $f(R)=x$). 
Second, for each agent $i\in N$ and for each of its types $R_i\in \mathcal{D}_i$, the strategy $\sigma ^{R_i}_i$ corresponding to its type $R_i$ must be obviously dominant. 
This means that whenever $i$ has to make a choice in $\Gamma$, it evaluates the consequence of playing according to $\sigma ^{R_i}_i$ in a pessimistic manner (anticipating the worst possible outcome) and evaluates the consequence of deviating to any other strategy $\sigma ^{\prime }_i$ optimistically (anticipating the best possible outcome).
Moreover, the pessimistic outcome associated with $\sigma ^{R_i}_i$ must be at least as good as the optimistic outcome associated with the deviation $\sigma^{\prime }_i$, according to $R_i$. 
Hence, whenever an agent has to play, the choice prescribed by the strategy corresponding to its type appears as evidently optimal; in other words, obviously dominant. 
In this case, we say that the extensive game form $\Gamma $ and the type-strategy profile $(\sigma_i^{R_i})_{R_i\in \mathcal{D}_i\,,\,i\in N}$ OSP-implement $f$.

The challenge in determining whether a social choice function $f$ is obviously strategy-proof lies in the requirement that its implementation in obviously dominant strategies must occur through an extensive game form, which is not provided, like in strategy-proofness, by a general revelation principle in the form of the direct revelation mechanism.
The primary difficulty lies in identifying, for each social choice function, the extensive game form $\Gamma$ used to OSP-implement $f$.

To propose intermediate OSP-implementability notions that require varying degrees of contingent reasoning, we must consider extensive game forms, which will be discussed in the following subsection.

\subsection{Extensive game forms}

Table 1 provides the basic notation for extensive game forms.

\begin{center}
\textsc{Table 1: Notation for Extensive Game Forms} \vspace{-0.3in} 
\begin{equation*}
\begin{tabular}{cccc}
&  &  &  \\ \hline\hline
{\small Name} &  & {\small Notation} & {\small Generic element} \\ \hline
\multicolumn{1}{l}{\small Players (or agents)} &  & $N$ & $i%
$ \\ 
\multicolumn{1}{l}{\small Outcomes (or alternatives)} &  & $ A$ & $%
x$ \\ 
\multicolumn{1}{l}{\small Histories} &  & $H$ & $ h$ \\ 
\multicolumn{1}{l}{\small Initial history} &  & ${h^0}$ &  \\ 
\multicolumn{1}{l}{\small Nodes} &  & $Z$ & $ z$ \\ 
\multicolumn{1}{l}{{\small Partial order on }$Z$} &  & $ \prec $ & 
\\ 
\multicolumn{1}{l}{\small Initial node} &  & $ z_{0}$ &  \\ 
\multicolumn{1}{l}{\small Terminal nodes} &  & $Z_{T}$ &  \\ 
\multicolumn{1}{l}{\small Non-terminal nodes} &  & $ Z_{NT}$ &  \\ 
\multicolumn{1}{l}{{\small Nodes where }$i${\small \ plays}} &  & $ Z
_{i}$ & $z_{i}$ \\ 
\multicolumn{1}{l}{{\small Information sets of player }$i$} &  & $\mathcal{I}
_{i}$ & $I_{i}$ \\ 
\multicolumn{1}{l}{{\small Choices (or actions) at }$z_i\in Z_{NT}$} &  & $Ch(z_i)
$ & $a_i$  \\ 
\multicolumn{1}{l}{{\small Outcome at }$z\in Z_{T}$} &  & $ g(z)$ & 
\\ \hline
\end{tabular}
\end{equation*}
\end{center}

\medskip

An extensive game form with a set of players $N$ and outcomes in $A$ (or simply, a \textit{game}) is a seven-tuple $\Gamma =(N,A,( Z,\prec ),\mathcal{Z},\mathcal{I},Ch,g)$, where $(Z,\prec )$ is a rooted tree, a graph with the properties that any two nodes in $Z$ are connected through a unique path and there exists a distinguished node $z_0\in Z_{NT}$, referred to as the root or initial node, such that $z_0\prec z$ for all $z\in Z\setminus \{z_0\}$. 
Alternatively, for every node $z\in Z\backslash \{z_{0}\}$, there exists a \emph{unique} node $z^{\prime}$ such that $z^{\prime}\prec z$ and no other node $z^{\prime \prime}\in Z_{NT}$ satisfies $z^{\prime }\prec z^{\prime \prime}\prec z$; this particular node $z^{\prime }$ is known as the immediate predecessor of $z$ and is denoted by $IP(z)$. 
In addition to the notation of Table 1, $\mathcal{Z}=\{Z_{1},\ldots ,Z_{n}\}$ represents the partition of $Z_{NT}$, where $z\in Z_i$ indicates that player $i$ plays at node $z$, $\mathcal{I}=\{\mathcal{I}_{1},\ldots ,\mathcal{I}_{n}\}$ represents the partition of information sets, where $z,z^{\prime} \in I_i\in \mathcal{I}_i$ indicates that player $i$ has to play at information set $I_i$ (\emph{i.e.}, $I_i\subseteq Z_i$) and does not know whether the game has reached node $z$ or $z^{\prime}$, and $Ch=\bigcup_{z\in Z_{NT}}Ch(z)$ is the collection of all available choices. 
Certainly, for each $z\in Z_{NT}$, there should be a one-to-one identification between $Ch(z)$ and the set of immediate followers of $z$, denoted as $IF(z)=\{z^{\prime }\in Z\mid IP(z^{\prime })=z \}$. 
Due to this correspondence, we often identify the choice made by agent $i$ at node $z\in Z_i$ with the subsequent node that immediately follows $z$. 
Furthermore, for each $I_i\in \mathcal{I}_i$ and any pair $z,z^{\prime }\in I_i$, $Ch(z)=Ch(z^{\prime })$ holds; namely, player $i$ at information set $I_i$ cannot distinguish between nodes $z$ and $z^{\prime }$ by observing the set of their available choices. 
Therefore, we denote the set of available choices at information set $I_i$ as $Ch(I_i)$, which is equivalent to $Ch(z)$ for any $z\in I_i$.
We denote $I_i^{\prime}\prec I_i$ if for every $z^{\prime }\in I_i^{\prime}$, there exists a node $z\in I_i$ such that $z^{\prime}\prec z$.
A history $h$ (of length $t$) is a sequence $z_{0},z_{1},\ldots ,z_{t}$ of $t+1$ nodes, starting at $z_{0}$ and ending at $z_t$, such that for all $m=1,\ldots ,t$, $z_{m-1}=IP(z_{m})$.
Each history $h=z_{0},\ldots ,z_{t}$ can be uniquely identified with the node $z_{t}$, and conversely, each node $z$ can be uniquely identified with the history $h=z_{0},\ldots ,z$. 
A history $h=z_0,\dots ,z_t$ is complete if $z_t\in Z_T$.
The outcome function $g:Z_T \rightarrow A$ assigns to each terminal node $z\in Z_T$ an outcome $g(z)\in A$.
Note that $\Gamma $ is not yet a game in extensive form because agents' preferences over outcomes (associated with terminal nodes) are not specified. 
However, given a game $\Gamma $ and a profile of preferences $R\in \mathcal{D}$ over $A$, the pair $(\Gamma ,R)$ defines a game in extensive form where each agent $i$ uses $R_{i}$ to evaluate pairs of outcomes, which are associated with pairs of terminal nodes.
Since $N$ and $A$ will remain constant throughout the paper, let $\mathcal{G}$ denote the class of all games with the set of players $N$ and outcomes in $A$. 
Henceforth, we will refer to $N$ as the set of agents and to $A$ as the set of alternatives.

Let $\Gamma \in \mathcal{G}$ and $i\in N$ be fixed. 
A (behavioral and pure) \textit{strategy} of $i$ in $\Gamma $ is a function $\sigma_{i}:Z_{i}\rightarrow Ch$, where for each $z\in Z_{i}$, $\sigma_{i}(z)\in Ch(z)$; namely, $\sigma _{i}$ selects one of $i$'s available choices at each node where $i$ must play. 
Additionally, $\sigma _{i}$ is $\mathcal{I}_{i}$-measurable: for any $I_i\in \mathcal{I}_{i}$ and any pair $z,z^{\prime }\in I_i$, $\sigma _{i}(z)=\sigma _{i}(z^{\prime })$. 
Therefore, we often use $\sigma _{i}(I_i)$ to denote the choice prescribed by $\sigma_i$ at all nodes in $I_i$.
Let $\Sigma _{i}$ represent the set of strategies available to agent $i$ in $\Gamma $. 
Then, a strategy profile $\sigma =(\sigma _{1},\ldots ,\sigma _{n})\in \Sigma := \Sigma _{1}\times \cdots \times \Sigma _{n}$ is an ordered list of strategies, one for each agent.
Let $z^{\Gamma }(z,\sigma )$ denote the terminal node reached in $\Gamma $ when agents commence playing at $z\in Z_{NT}$ according to $\sigma \in \Sigma $. 
Given $\sigma \in \Sigma$ and $S\subseteq N$, $\sigma_S=(\sigma_i)_{i\in S}$ represents the strategy profile of agents in $S$.

In the context of a given game $\Gamma$ and a domain $\mathcal{D}$, a \textit{type-strategy} profile $(\sigma^{R_i}_i)_{R_i\in \mathcal{D}_i\,,\,i\in N}$ specifies, for each agent $i\in N$ and preference $R_i\in \mathcal{D}_i$, a behavioral strategy $\sigma ^{R_i}_i\in \Sigma _i$ of $i$ in $\Gamma $. 
We denote the strategy profile $(\sigma_1 ^{R_1}, \dots ,\sigma_n^{R_n})\in \Sigma$ as $\sigma ^R$.

Let $f:\mathcal{D}\rightarrow A$ be a social choice function and let $\Gamma $ be an extensive game form representation of its normal game form $(N,\mathcal{D},f)$.
Namely, each agent $i$ only plays once in $\Gamma$ by choosing one preference in $\mathcal{D}_i$, without knowing the choices made by the other agents.
For each $R_i$, let $\sigma^{R_i}_i$ be the truth-telling strategy that prescribes choosing $R_i$ at the unique information set of agent $i$ in $\Gamma $.
The outcome $g(z^\Gamma (z_0,\sigma^R))$ associated to the terminal node induced by the strategy profile $\sigma^R$ is $f(R)$.
It is immediate to see that if $f:\mathcal{D}\rightarrow A$ is strategy-proof, then $\sigma ^{R_i}_i$ is a weakly dominant strategy in $\Gamma $ for all $i\in N$ and $R_i\in \mathcal{D}_i$; namely, for all $\sigma _{-i}\in \Sigma_{-i}$ and $\sigma ^{\prime }_i\in \Sigma _i$, 
\begin{equation*}
g(z^\Gamma (z_0,(\sigma ^{R_i}_i,\sigma _{-i})))\,R_i\,g(z^\Gamma (z_0,(\sigma ^{\prime}_i,\sigma _{-i}))).
\end{equation*}

\section{Obvious strategy-proofness with respect to a partition}

\subsection{Definition and example}

To define obvious strategy-proofness with respect to a partition of agents $\mathcal{S}=\{S_1,\dots ,S_K\}$, where $1\leq K\leq n$, we introduce several necessary concepts.

Fix a game $\Gamma \in \mathcal{G}$ and a subset of agents $S\subseteq N$.

A history $h = z_0, \ldots, z_t$ (or node $z_t$) is \textit{compatible with} $\sigma_S$ if, for all $z_{t^\prime} \in Z_i$ such that $0 \leq t^\prime < t$ and $i\in S$, $\sigma_i(z_{t^\prime}) = z_{t^\prime + 1}$ holds. 
In other words, a history $h = z_0, \ldots, z_t$ is compatible with $\sigma_S$ if, whenever an agent $i\in S$ must play at a node $z_{t^\prime}$ on the path from $z_0$ to $z_t$, the choice made by agent $i$ according to $\sigma_i$ results in the node $z_{t^\prime + 1}$.
It's important to note that the compatibility of $h = z_0, \dots, z_t$ with $\sigma_S$ does not preclude the possibility of an agent not in $S$ playing along the history toward $z_t$. 
Specifically, it's possible to have $z_{t^\prime} \in Z_i$ for some $0 \leq t^\prime < t$ and $i \notin S$.
Given $\sigma _S$ and an agent $i \in S$, along with an alternative strategy $\sigma_ {i}^{\prime } \in \Sigma_{i}\setminus\{\sigma_i\}$, an earliest point of departure for $\sigma _{S}$ and $\sigma_{i}^{\prime }$ consists of a set of nodes compatible with $\sigma_S$ within an information set $I_i$. 
These nodes are characterized by the property that $\sigma_{i} $ and $\sigma _{i}^{\prime }$ prescribe different actions at each of them but identical actions at all previous information sets encountered along each of their respective paths.

\begin{definition}\label{DefEPoD}
Given $\sigma_S$, $i\in S$, $\sigma_{i}^{\prime }\in \Sigma_{i}\setminus\{\sigma_i\}$, and $I_{i}\in \mathcal{I}_{i}$, we define a set of nodes $z\in I_{i}$ that are compatible with $\sigma _{S}$, denoted by $I_i(\sigma_S,\sigma_i^{\prime})$, as an \emph{earliest point of departure for $\sigma _{S}$ and $\sigma _{i}^{\prime }$} if

\noindent (i) $\sigma _{i}(I_{i})\neq \sigma _{i}^{\prime }(I_{i})$,

\noindent (ii) $\sigma _{i}(I_{i}^{\prime })=\sigma _{i}^{\prime}(I_{i}^{\prime })$ for all $I_{i}^{\prime }\in \mathcal{I}_i$ such that $I_{i}^{\prime } \prec I_{i}$.
    
\end{definition}

Two key observations can be made. 
First, an earliest point of departure constitutes a subset of an agent's information set. 
Second, it is defined relative to a joint strategy $\sigma _S$ employed by agents in $S$, of which $i$ is a member, and to an alternative strategy $\sigma ^{\prime }_i$ distinct from the strategy $\sigma _i$ specified within $\sigma _S$. 
To illustrate this concept, let's consider the game $\Gamma$ depicted in Figure 1 below. A full description of this game will be provided later on.
Let $S=\{1,2\}$, $(\sigma _1,\sigma _2)$ and $\sigma ^{\prime}_2$ be such that $\sigma_1(z_0)=y$, $\sigma_2(I_2)=y$ and $\sigma ^{\prime}_2(I_2)=x$.
Then, the earliest point of departure for $(\sigma_1,\sigma_2)$ and $\sigma_2^{\prime}$ is $I_2((\sigma_1,\sigma_2),\sigma^{\prime}_2)=\{z_1\}\subsetneq I_2$. 
Furthermore, earliest points of departure may be strict subsets of information sets because the strategies of all agents in $S$ except $i$ have been fixed, thereby excluding nodes within the same information set.\footnote{Note that if we take $S=\{2\}$, then $I_2(\sigma_2,\sigma_2')=I_2$.\smallskip} 

Let $\sigma _{S}$ and $\sigma _{i}^{\prime }$ be given. 
We denote the set of earliest points of departure for $\sigma _{S}$ and $\sigma _{i}^{\prime }$ by $\alpha (\sigma _{S},\sigma _{i}^{\prime })$.

Given the partition $\mathcal{S}$ of $N$ and an agent $i\in N$, let $S^i\in \mathcal{S}$ denote the element of $\mathcal{S}$ containing $i$.
Considering $\sigma_{S^i}$ and $\sigma ^{\prime }_i$, we define $o(\sigma_{S^i},\sigma^{\prime }_i)$ and $o^{\prime }(\sigma_{S^i},\sigma^{\prime }_i)$ as the sets of options respectively remaining after $\sigma_i$ and $\sigma^{\prime }_i$ at the earliest point of departure $I_{i}(\sigma _{S^i},\sigma_i^{\prime})$; namely,\footnote{It is important to note that $o(\sigma_{S^i},\sigma ^{\prime }_i)$ and $o^{\prime }(\sigma_{S^i},\sigma^{\prime }_i)$ depend on the choice of $I_{i}(\sigma _{S^i},\sigma_i^{\prime})$, although this aspect is not explicitly reflected in the notation. 
However, in all cases, the context will make it clear which earliest point of departure is being referred to, thus avoiding confusion.\smallskip}
\begin{equation*}
o(\sigma_{S^i},\sigma ^{\prime }_i)=\{x\in A\mid \exists \overline{\sigma }_{-S^i}\in \Sigma _{-S^i} \text{ and }z\in I_{i}(\sigma_{S^i},\sigma_i^{\prime})\text{ s.t. }x=g(z^{\Gamma }(z,(\sigma _{i},\sigma_{S^i\setminus \{i\}},\overline{\sigma }_{-S^i})))\}
\end{equation*}
and 
\begin{equation*}
o^{\prime }(\sigma_{S^i},\sigma^{\prime }_i)=\{y\in A\mid \exists \overline{\sigma }_{-S^i}\in \Sigma _{-S^i} \text{ and }z\in I_{i}(\sigma_{S^i},\sigma_i^{\prime})\text{ s.t. }y=g(z^{\Gamma }(z,(\sigma^{\prime}_{i},\sigma _{S^i\setminus \{i\}},\overline{\sigma }_{-S^i})))\}.
\end{equation*}

With the necessary concepts in place, we can now proceed to define the notion of obviously dominant strategy with respect to a partition of agents $\mathcal{S}$, given a game $\Gamma$ and a domain of preferences $\mathcal{D}$. \medskip

\begin{definition}\label{DefODwS}

A strategy $\sigma _{i}$ is \emph{obviously dominant with respect to a partition $\mathcal{S}$ in $\Gamma$ for agent $i$ with $R_{i}\in \mathcal{D}_i$} if, for all $\sigma _{S^i\setminus \{i\}}\in \Sigma_{S^i\setminus \{i\}}$, all $\sigma _{i}^{\prime }\neq \sigma _{i}$ and all $I_i(\sigma _{S^i},\sigma_i^{\prime})\in \alpha (\sigma_{S^i},\sigma_{i}^{\prime })$, the following holds:
For all $x\in o(\sigma_{S^i},\sigma^{\prime }_i)$ and all $y\in o^{\prime }(\sigma_{S^i},\sigma^{\prime }_i)$, 
\begin{eqnarray*}
x\,R_i\,y.
\end{eqnarray*}
\end{definition}
In words, when $i$ follows strategy $\sigma_i$ given any $\sigma_{S^i}$, the least desirable alternative that $i$ can achieve is at least as favorable, according to preference relation $R_i$, as the best alternative attainable by $i$ if it deviates from $\sigma_i$ to $\sigma ^{\prime }_i$. In this sense, $\sigma _i$ is unquestionably superior.

\begin{definition}\label{DefOSPwS}
A social choice function $f:\mathcal{D}\rightarrow A$ is \emph{obviously strategy-proof (OSP) with respect to a partition} $\mathcal{S}$ if there exist an extensive game form $\Gamma \in \mathcal{G}$ and a type-strategy profile $(\sigma_i^{R_i})_{R_i\in \mathcal{D}_i\,,\,i\in N}$ for $\Gamma $ such that, for each $R\in \mathcal{D}$:

\noindent (i) $f(R)=g(z^{\Gamma }(z_{0},\sigma ^{R}))$ and

\noindent (ii) for all $i\in N$, $\sigma_{i}^{R_{i}}$ is obviously dominant with respect to $\mathcal{S}$ in $\Gamma $ for $i$ with $R_{i}$.
\end{definition}

When (i) holds we say that $\Gamma $ and $(\sigma_i^{R_i})_{R_i\in \mathcal{D}_i\,,\,i\in N}$ induce $f$. When (i) and (ii) hold we say that $\Gamma $ and $(\sigma_i^{R_i})_{R_i\in \mathcal{D}_i\,,\,i\in N}$ OSP-\textit{implement} $f$ with respect to $\mathcal{S}$. 
We often omit the explicit reference to the type-strategy profile and simply say that $\Gamma $ OSP-implements $f$.

\begin{remark}\label{RemK=1,K=n}
Let $f:\mathcal{D}\rightarrow A$ be a social choice function. 
Then,

\noindent \emph{(R1.1)} $f$ is OSP with respect to $\mathcal{S}=\{N\}$ if and only if $f$ is SP.

\noindent \emph{(R1.2)} $f$ is OSP with respect to $\mathcal{S}=\{\{1\},\dots ,\{n\}\}$ if and only if $f$ is OSP.\footnote{Namely, Li (2017)'s definition of obvious strategy-proofness for our setting corresponds to OSP with respect to the finest partition of $N$.
This means that each agent $i$, when evaluating the consequences of $\sigma_i$ and $\sigma_i^\prime$ at its earliest points of departure, considers only its own strategy $\sigma_i$ as given.\smallskip}
\end{remark}

Example 1 illustrates the notion of obvious strategy-proofness with respect to a partition.\medskip

\noindent \textbf{Example 1.} Let $N=\{1,2,3,4,5\}$ be the set of agents, let $\mathcal{S^\ast}=\{\{1,2\},\{3\},\{4,5\}\}$ be the partition, and let $A=\{x,y\}$ be the set of alternatives. 
For each $i\in N$, let $\mathcal{D}_i=\mathcal{P}=\{P^x,P^y\}$ be the domain of the two strict preferences over $A$, where $x\,P^x\,y$ and $y\,P^y\,x$ (\textit{i.e.}, $x=t(P^x)$ and $y=t(P^y)$). 
When it does not lead to any confusion, we will refer to $P^x$ and $P^y$ only by their preferred alternatives $x$ and $y$, respectively. 
Define the social choice function $f:\mathcal{P}^N\rightarrow \{x,y\}$ as follows: 
For each $P\in \mathcal{P}^N$, $f(P)=x$ if (i) $t(P_1)=t(P_2)=x$, or (ii) $t(P_1)=t(P_3)=x$ or (iii) $t(P_2)=t(P_4)=t(P_5)=x$ hold; otherwise, $f(P)=y$.\footnote{
This is a particular instance of an extended majority voting rule that we shall define later through a family of minimal winning coalitions for $x$, $\mathcal{C}_m^x$. 
The family contains those subsets of agents that can impose $x$ whenever all their members declare $x$ as their top alternative; in this case, $\mathcal{C}^x_m=\{\{1,2\},\{1,3\},\{2,4,5\}\}$. By Arribillaga, Massó and Neme (2020), this voting rule is not obviously strategy-proof.\smallskip}

\begin{center}
\unitlength=1.5mm 
\begin{picture}(62,56)(0,0)
{\scriptsize

   \put(34,50){\circle*{1}}
   \put(14,40){\circle*{1}}
   \put(54,40){\circle*{1}}
   \put(6,30){\circle*{1}}
   \put(22,30){\circle*{1}}
   \put(48,30){\circle*{1}}
   \put(60,30){\circle*{1}}
   \put(12,20){\circle*{1}}
   \put(32,20){\circle*{1}}
   \put(42,20){\circle*{1}}
   \put(54,20){\circle*{1}}
   \put(8,10){\circle*{1}}
   \put(16,10){\circle*{1}}
   \put(28,10){\circle*{1}}
   \put(36,10){\circle*{1}} 

   \put(33.3,47.5){\mbox{$z_0$}}
   \put(10.5,39.5){\mbox{$z_1$}}
   \put(55.5,39.5){\mbox{$z_2$}}
   \put(33.3,41){\mbox{$I_2$}}
   \put(18.5,29.5){\mbox{$z_4$}}  
   \put(49.5,29.5){\mbox{$z_3$}}
   \put(8.5,19.5){\mbox{$z_5$}}
   \put(33.5,19.5){\mbox{$z_6$}}
   \put(21.3,21){\mbox{$I_5$}}
     
   \put(34,50){\line(-2,-1){20}}
   \put(34,50){\line(2,-1){20}}
   \put(14,40){\line(-4,-5){8}}
   \put(14,40){\line(4,-5){8}}
   \put(54,40){\line(-3,-5){6}}
   \put(54,40){\line(3,-5){6}}
   \put(22,30){\line(-1,-1){10}}
   \put(22,30){\line(1,-1){10}}
   \put(48,30){\line(-3,-5){6}}
   \put(48,30){\line(3,-5){6}}
   \put(12,20){\line(-2,-5){4}}
   \put(12,20){\line(2,-5){4}}
   \put(32,20){\line(-2,-5){4}}
   \put(32,20){\line(2,-5){4}}
   
   \put(33.5,51.2){\mbox{\textbf{1}}}
   \put(13.5,41.2){\mbox{\textbf{2}}}
   \put(54.5,41.2){\mbox{\textbf{2}}}
   \put(22.5,31.2){\mbox{\textbf{4}}}
   \put(45.5,31.2){\mbox{\textbf{3}}}
   \put(11.5,21.2){\mbox{\textbf{5}}}
   \put(32.5,21.2){\mbox{\textbf{5}}}
   
   \put(23,46){\mbox{$y$}}
   \put(44,46){\mbox{$x$}}   
   \put(9,36){\mbox{$y$}}
   \put(18,36){\mbox{$x$}}   
   \put(50,36){\mbox{$y$}}   
   \put(57,36){\mbox{$x$}}
   \put(16,26){\mbox{$y$}}
   \put(27,26){\mbox{$x$}}
   \put(44,26){\mbox{$y$}}
   \put(51,26){\mbox{$x$}}
   \put(9,16){\mbox{$y$}}
   \put(14,16){\mbox{$x$}}
   \put(29,16){\mbox{$y$}}
   \put(34,16){\mbox{$x$}}

   \put(5.5,27.5){\mbox{$y$}}
   \put(7.5,7.5){\mbox{$y$}}
   \put(15.5,7.5){\mbox{$y$}}
   \put(27.5,7.5){\mbox{$y$}}
   \put(35.5,7.5){\mbox{$x$}}
   \put(41.5,17.5){\mbox{$y$}}
   \put(53.5,17.5){\mbox{$x$}}
   \put(59.5,27.5){\mbox{$x$}}

   \multiput(14.5,40)(2,0){20}
   {\line(1,0){0.5}}
      
   \multiput(12.5,20)(2,0){10}
   {\line(1,0){0.5}}
 
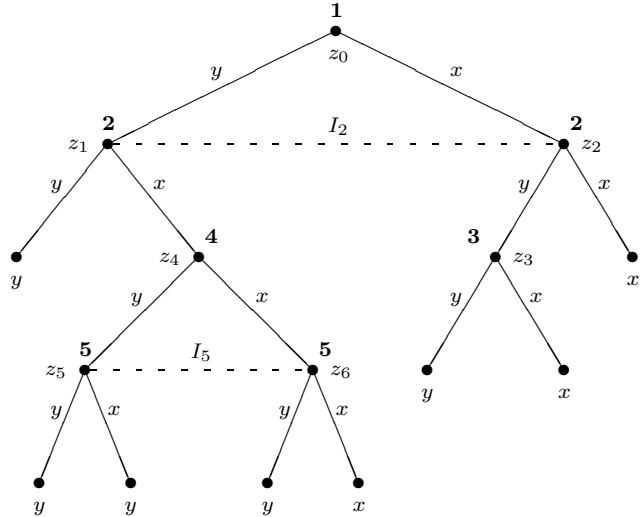
   \put(6,0){Figure 1: An extensive game form $\Gamma $ that illustrates Definition 3}}

\end{picture}
\end{center}

Figure 1 above depicts the extensive game form $\Gamma \in \mathcal{G}$, where agents play only once, information sets of agents 1, 3 and 4 contain a unique node ($z_0$, $z_3$ and $z_4$, respectively), and agents 2 and 5 have an information set with two nodes ($I_2=\{z_1,z_2\}$ and $I_5=\{z_5,z_6\}$, respectively) and, at each $z\in Z_{NT}$, $Ch(z)=\{x,y\}$.

For agent $i\in N$ with preference $P_i\in \mathcal{P}$, define the truth-telling strategy $\sigma_i^{P_i}$ by setting $\sigma_i^{P_i}(z)=t(P_i)$ for $z\in Z_i$.
It is easy to check that the particular social choice function $f$, defined above, is induced by $\Gamma $ and $(\sigma_i^{P_i})_{P_i\in \mathcal{P},\,i\in N}$.
To complete the verification that $f$ is OSP with respect to $\mathcal{S}^\ast$, we check that, for each $i\in N$ and each $P_i\in \mathcal{P}$, $\sigma^{P_i}_i$ is obviously dominant with respect to $\mathcal{S}^\ast=\{\{1,2\},\{3\},\{4,5\}\}$ in $\Gamma $ for $i$ with $P_i$.

Consider coalition $S_1^\ast=\{1,2\}$ and agent $1$.

Assume $x\,P_1\,y$ (\textit{i.e.}, $P_1=P^x$). 
Then, agent $1$'s truth-telling strategy is $\sigma_1^{P_1}(z_0)=x$ and let $\sigma_1^{\prime}(z_0)=y$ be agent $1$'s deviating strategy. 
For any $\sigma_2\in\Sigma_2$, write $\sigma_{S_1^\ast}=(\sigma^{P_1}_1,\sigma_2)$.
Fix $\sigma_2(I_2)=x$.
Hence, $\alpha (\sigma _{S_1^\ast},\sigma _{1}^{\prime })=\{I_1(\sigma_{S_1^\ast}, \sigma^\prime _1)\}$ and $I_1(\sigma_{S_1^\ast}, \sigma^\prime _1)=\{z_0\}$, and so $o(\sigma_{S_1^\ast},\sigma ^{\prime}_1)=\{x\}$ and $o^\prime (\sigma_{S_1^\ast},\sigma ^{\prime }_1)=\{x,y\}$. 
Then, $x$ is the worst (and unique) alternative of playing according to the truth-telling strategy $\sigma_1^{P_1}(z_0)=x$, which is weakly preferred to $x$, the best possible alternative of playing according to the deviating strategy $\sigma_1^\prime (z_0)=y$. 
Fix $\sigma_2(I_2)=y$. 
Hence, $\alpha (\sigma _{S_1^\ast},\sigma _{1}^{\prime })=\{I_1(\sigma_{S_1^\ast}, \sigma^\prime _1)\}$ and $I_1(\sigma_{S_1^\ast},\sigma^\prime _1)=\{z_0\}$ and so $o(\sigma_{S_1^\ast},\sigma ^{\prime }_1)=\{x,y\}$ and $o^{\prime }(\sigma_{S_1^\ast},\sigma ^{\prime }_1)=\{y\}$. 
Then, $y$ is the worst possible alternative of playing according to the truth-telling strategy $\sigma_1^{P_1}(z_0)=x$, which is weakly preferred to $y$, the best (and unique) alternative of playing according to the deviating strategy $\sigma_1^\prime (z_0)=y$.

Assume $y\,P_1\,x$ (\textit{i.e.}, $P_1=P^y$). 
Then, agent $1$'s truth-telling strategy is $\sigma_1^{P_1}(z_0)=y$ and let $\sigma_1^\prime (z_0)=x$ be agent $1$'s deviating strategy.
For any $\sigma_2\in\Sigma_2$, write $\sigma_{S_1^\ast}=(\sigma^{P_1}_1,\sigma_2)$.
Fix $\sigma_2(I_2)=x$. 
Hence, $\alpha (\sigma _{S_1^\ast},\sigma _{1}^{\prime })=\{I_1(\sigma_{S_1^\ast}, \sigma^\prime _1)\}$ and $I_1(\sigma_{S_1^\ast},\sigma^\prime _1)=\{z_0\}$, and so $o(\sigma_{S_1^\ast},\sigma ^{\prime}_1)=\{x,y\}$ and $o^{\prime }(\sigma_{S_1^\ast},\sigma ^{\prime}_1)=\{x\}$.
Then, $x$ is the worst possible alternative of playing according to the truth-telling strategy $\sigma_1^{P_1}(z_0)=y$, which is weakly preferred to $x$, the best (and unique) alternative of playing according to the deviating strategy $\sigma_1^\prime (z_0)=x$. 
Fix $\sigma_2(I_2)=y$. 
Hence, $\alpha (\sigma _{S_1^\ast},\sigma _{1}^{\prime })=\{I_1(\sigma_{S_1^\ast}, \sigma^\prime _1)\}$ and $I_1(\sigma_{S_1^\ast},\sigma^\prime _1)=\{z_0\}$ and so $o(\sigma_{S_1^\ast},\sigma ^{\prime }_1)=\{y\}$ and $o^{\prime}(\sigma_{S_1^\ast},\sigma ^{\prime }_1)=\{x,y\}$. 
Then, $y$ is the worst (and unique) alternative of playing according to the truth-telling strategy $\sigma_1^{P_1}(z_0)=y$, which is weakly preferred to $y$, the best possible alternative of playing according to the deviating strategy $\sigma_1^\prime(z_0)=x$.

Consider now agent $2$.

Assume $x\,P_2\,y$ (\textit{i.e.}, $P_2=P^x$).
Then, agent $2$'s truth-telling strategy is $\sigma_2^{P_2}(I_2)=x$ and let $\sigma_2^\prime (I_2)=y$ be agent $2$'s deviating strategy. 
For any $\sigma_1\in\Sigma_1$, write $\sigma_{S_1^\ast}=(\sigma_1,\sigma^{P_2}_2)$.
Fix $\sigma_1(z_0)=x$. 
Hence, $\alpha (\sigma _{S_1^\ast},\sigma _{2}^{\prime })=\{I_2(\sigma_{S_1^\ast}, \sigma^\prime _2)\}$ and $I_2(\sigma_{\mathcal{S}^\ast _1},\sigma^\prime_2)=\{z_2\}$, and so $o(\sigma_{S^{\ast }_1},\sigma^{\prime }_2)=\{x\}$ and $o^{\prime }(\sigma_{S^{\ast }_1},\sigma ^{\prime}_2)=\{x,y\}$. 
Then, $x$ is the worst (and unique) alternative of playing according to the truth-telling strategy $\sigma_2^{P_2}(I_2)=x$, which is weakly preferred to $x$, the best possible alternative of playing according to the deviating strategy $\sigma_2^\prime (I_2)=y$. Fix $\sigma_1(z_0)=y$.
Hence, $\alpha (\sigma _{S_1^\ast},\sigma _{2}^{\prime })=\{I_2(\sigma_{S_1^\ast}, \sigma^\prime _2)\}$ and $I_2(\sigma_{\mathcal{S}^\ast _1},\sigma^\prime_2)=\{z_1\}$, and so $o(\sigma_{S^{\ast }_1},\sigma ^{\prime }_2)=\{x,y\}$ and $o^{\prime}(\sigma_{S^{\ast }_1},\sigma ^{\prime }_2)=\{y\}$. 
Then, $y$ is the worst possible alternative of playing according the truth-telling strategy $\sigma_2^{P_2}(I_2)=x$, which is weakly preferred to $y$, the best (and unique) alternative of playing according to the deviating strategy $\sigma_2^\prime (I_2)=y$.

Assume $y\,P_2\,x$  (\textit{i.e.}, $P_2=P^y$). 
Then, agent $2$'s truth-telling strategy is $\sigma_2^{P_2}(I_2)=y$ and let $\sigma_2^\prime (I_2)=x$ be agent $2$'s deviating strategy. 
For any $\sigma_1\in\Sigma_1$, write $\sigma_{S_1^\ast}=(\sigma_1,\sigma^{P_2}_2)$.
Fix $\sigma_1(z_0)=x$. 
Hence, $\alpha (\sigma _{S_1^\ast},\sigma _{2}^{\prime })=\{I_2(\sigma_{S_1^\ast}, \sigma^\prime _2)\}$ and $I_2(\sigma_{\mathcal{S}^\ast _1},\sigma^\prime_2)=\{z_2\}$, and so $o(\sigma_{S^{\ast }_1},\sigma^{\prime }_2)=\{x,y\}$ and $o^{\prime }(\sigma_{S^{\ast }_1},\sigma ^{\prime}_2)=\{x\}$. 
Then, $x$ is the worst possible alternative of playing according to the truth-telling strategy $\sigma_2^{P_2}(I_2)=y$, which is weakly preferred to $x$, the best (and unique) alternative of playing according to the deviating strategy $\sigma_2^\prime (I_2)=x$. 
Fix $\sigma_1(z_0)=y$. 
Hence, $\alpha (\sigma _{S_1^\ast},\sigma _{2}^{\prime })=\{I_2(\sigma_{S_1^\ast}, \sigma^\prime _2)\}$ and $I_2(\sigma_{\mathcal{S}^\ast_1},\sigma^\prime_2)=\{z_1\}$, and so $o(\sigma_{S^{\ast }_1},\sigma^{\prime }_2)=\{y\}$ and $o^{\prime }(\sigma_{S^{\ast }_1},\sigma ^{\prime}_2)=\{x,y\}$. 
Then, $y$ is the worst (and unique) possible alternative of playing according to the truth-telling strategy $\sigma_2^{P_2}(I_2)=y$, which is weakly preferred to $y$, the best possible alternative of playing according to the deviating strategy $\sigma_2^\prime (I_2)=x$.

Therefore, truth-telling is obviously dominant with respect to $\mathcal{S}^\ast$ in $\Gamma $ for agents $1$ and $2$ with each of the two preferences.

Consider coalition $S_2^\ast=\{3\}$. 
For any $P_3\in \mathcal{P}$ and deviating strategy $\sigma ^{\prime }_3$, $\alpha (\sigma^{P_3}_3,\sigma^{\prime}_3)=\{I_3(\sigma^{P_3}_3,\sigma^{\prime}_3)\}$ and $I_3(\sigma^{P_3}_3,\sigma^{\prime}_3)=\{z_3\}$ hold, and so $o(\sigma_{S^{\ast }_3},\sigma ^{\prime }_3)=t(P_3)$, and $o^{\prime }(\sigma_{S^{\ast }_3},\sigma ^{\prime }_3)\neq t(P_3)$ hold.
Then, $t(P_3)$ is the worst (and unique) possible alternative of playing according to the truth-telling strategy, which is strictly preferred to $\sigma ^{\prime }_3(I_3)\neq t(P_3)$, the best possible alternative of playing according to the deviating strategy.

Therefore, truth-telling is obviously dominant with respect to $\mathcal{S}^\ast$ in $\Gamma $ for agent $3$ with each of the two preferences.

Consider coalition $S_3^\ast=\{4,5\}$ and agent $4$.

Assume $x\,P_4\,y$ (\textit{i.e.}, $P_4=P^x$).
Then, agent $4$'s truth-telling strategy is $\sigma_4^{P_4}(z_4)=x$ and let $\sigma_4^{\prime}(z_4)=y$ be agent $4$'s deviating strategy. 
For any $\sigma_5\in\Sigma_5$, write $\sigma_{S_3^\ast}=(\sigma^{P_4}_4,\sigma_5)$.
Fix $\sigma_5(I_5)=x$. 
Hence, $\alpha (\sigma _{S_3^\ast},\sigma _{4}^{\prime })=\{I_4(\sigma_{S_3^\ast}, \sigma^\prime _4)\}$ and  $I_4(\sigma_{S_3^\ast},\sigma^\prime _4)=\{z_4\}$, and so $o(\sigma_{S^{\ast }_3},\sigma ^{\prime}_4)=\{x\}$ and $o^\prime (\sigma_{S^{\ast }_3},\sigma ^{\prime }_4)=\{y\}$.
Then, $x$ is the worst (and unique) alternative of playing according to the truth-telling strategy $\sigma_4^{P_4}(z_4)=x$, which is strictly preferred to $y$, the best (and unique) alternative of playing according to the deviating strategy $\sigma_4^\prime (z_4)=y$. 
Fix $\sigma_5(I_2)=y$. 
Hence, $\alpha (\sigma _{S_3^\ast},\sigma _{4}^{\prime })=\{I_4(\sigma_{S_3^\ast}, \sigma^\prime _4)\}$ and $I_4(\sigma_{S_3^\ast},\sigma^\prime _4)=\{z_4\}$ and so $o(\sigma_{S^{\ast}_3},\sigma ^{\prime }_4)=\{y\}$ and $o^{\prime }(\sigma_{S^{\ast}_3},\sigma ^{\prime }_4)=\{y\}$.
Then, $y$ is the worst (and unique) alternative of playing according to the truth-telling strategy $\sigma_4^{P_4}(z_4)=y$, which is weakly preferred to $y$, the best (and unique) alternative of playing according to the deviating strategy $\sigma_4^\prime (z_4)=x$.

Assume $y\,P_4\,x$ (\textit{i.e.}, $P_4=P^y$).
Then, agent $4$'s truth-telling strategy is $\sigma_4^{P_4}(z_4)=y$ and let $\sigma_4^{\prime}(z_4)=x$ be agent $4$'s deviating strategy.
For any $\sigma_5\in\Sigma_5$, write $\sigma_{S_3^\ast}=(\sigma^{P_4}_4,\sigma_5)$.
Fix $\sigma_5(I_5)=x$. 
Hence, $\alpha (\sigma _{S_3^\ast},\sigma _{4}^{\prime })=\{I_4(\sigma_{S_3^\ast}, \sigma^\prime _4)\}$ and $I_4(\sigma_{S_3^\ast},\sigma^\prime _4)=\{z_4\}$, and so $o(\sigma_{S^{\ast }_3},\sigma ^{\prime}_4)=\{y\}$ and $o^\prime (\sigma_{S^{\ast }_3},\sigma ^{\prime }_4)=\{x\}$.
Then, $y$ is the worst (and unique) alternative of playing according to the truth-telling strategy $\sigma_4^{P_4}(z_4)=y$, which is strictly preferred to $y$, the best (and unique) alternative of playing according to the deviating strategy $\sigma_4^\prime (z_4)=x$. 
Fix $\sigma_5(I_2)=y$. 
Hence, $\alpha (\sigma _{S_3^\ast},\sigma _{4}^{\prime })=\{I_4(\sigma_{S_3^\ast}, \sigma^\prime _4)\}$ and $I_4(\sigma_{S_3^\ast},\sigma^\prime _4)=\{z_4\}$ and so $o(\sigma_{S^{\ast}_3},\sigma ^{\prime }_4)=\{y\}$ and $o^{\prime }(\sigma_{S^{\ast}_3},\sigma ^{\prime }_4)=\{y\}$.
Then, $y$ is the worst (and unique) alternative of playing according to the truth-telling strategy $\sigma_4^{P_4}(z_4)=y$, which is weakly preferred to $y$, the best (and unique) alternative of playing according to the deviating strategy $\sigma_4^\prime (z_4)=x$.

Consider now agent $5$.

Assume $x\,P_5\,y$  (\textit{i.e.}, $P_5=P^x$).
Then, agent $5$'s truth-telling strategy is $\sigma_5^{P_5}(I_5)=x$ and let $\sigma_5^{\prime}(I_5)=y$ be agent $5$'s deviating strategy. 
For any $\sigma_4\in\Sigma_4$, write $\sigma_{S_3^\ast}=(\sigma_4,\sigma^{P_5}_5)$.
Fix $\sigma_4(x_4)=x$. 
Hence, $\alpha (\sigma _{S_3^\ast},\sigma _{5}^{\prime })=\{I_5(\sigma_{S_3^\ast}, \sigma^\prime _5)\}$ and $I_5(\sigma_{S_3^\ast},\sigma^\prime _5)=\{z_6\}$, and so $o(\sigma_{S^{\ast }_3},\sigma ^{\prime}_5)=\{x\}$ and $o^\prime (\sigma_{S^{\ast }_3},\sigma ^{\prime }_5)=\{y\}$.
Then, $x$ is the worst (and unique) alternative of playing according to the truth-telling strategy $\sigma_5^{P_5}(I_5)=x$, which is strictly preferred to $y$, the best (and unique) alternative of playing according to the deviating strategy $\sigma_5^\prime (I_5)=y$. 
Fix $\sigma_4(z_4)=y$. 
Hence, $\alpha (\sigma _{S_3^\ast},\sigma _{5}^{\prime })=\{I_5(\sigma_{S_3^\ast}, \sigma^\prime _5)\}$ and $I_5(\sigma_{S_3^\ast},\sigma^\prime _5)=\{z_5\}$ and so $o(\sigma_{S^{\ast}_3},\sigma ^{\prime }_5)=\{y\}$ and $o^{\prime }(\sigma_{S^{\ast}_3},\sigma ^{\prime }_5)=\{y\}$. 
Then, $y$ is the worst (and unique) alternative of playing according to the truth-telling strategy $\sigma_5^{P_5}(I_5)=y$, which is weakly preferred to $y$, the best (and unique) alternative of playing according to the deviating strategy $\sigma_5^\prime (I_5)=y$.

Assume $y\,P_5\,x$ (\textit{i.e.}, $P_5=P^y$).
Then, agent $5$'s truth-telling strategy is $\sigma_5^{P_5}(I_5)=y$ and let $\sigma_5^{\prime}(I_5)=x$ be agent $5$'s deviating strategy.
For any $\sigma_4\in\Sigma_4$, write $\sigma_{S_3^\ast}=(\sigma_4,\sigma^{P_5}_5)$.
Fix $\sigma_4(x_4)=x$. 
Fix $\sigma_4(z_4)=x$. 
Hence, $\alpha (\sigma _{S_3^\ast},\sigma _{5}^{\prime })=\{I_5(\sigma_{S_3^\ast}, \sigma^\prime _5)\}$ and $I_5(\sigma_{S_3^\ast},\sigma^\prime _5)=\{z_6\}$, and so $o(\sigma_{S^{\ast }_3},\sigma ^{\prime}_5)=\{y\}$ and $o^\prime (\sigma_{S^{\ast }_3},\sigma ^{\prime }_5)=\{x\}$.
Then, $y$ is the worst (and unique) alternative of playing according to the truth-telling strategy $\sigma_5^{P_5}(I_5)=y$, which is strictly preferred to $x$, the best (and unique) alternative of playing according to the deviating strategy $\sigma_5^\prime (I_5)=x$. 
Fix $\sigma_4(z_4)=y$. 
Hence, $\alpha (\sigma _{S_3^\ast},\sigma _{5}^{\prime })=\{I_5(\sigma_{S_3^\ast}, \sigma^\prime _5)\}$ and $I_5(\sigma_{S_3^\ast},\sigma^\prime _5)=\{z_5\}$ and so $o(\sigma_{S^{\ast}_3},\sigma ^{\prime }_5)=\{y\}$ and $o^{\prime }(\sigma_{S^{\ast}_3},\sigma ^{\prime }_5)=\{y\}$.
Then, $y$ is the worst (and unique) alternative of playing according to the truth-telling strategy $\sigma_5^{P_5}(I_5)=y$, which is weakly preferred to $y$, the best (and unique) alternative of playing according to the deviating strategy $\sigma_5^\prime (I_5)=y$.

Therefore, truth-telling is obviously dominant with respect to $\mathcal{S}^\ast $ in $\Gamma $ for agents $4$ and $5$ with each of the two preferences. 
Thus, $\Gamma $ and $(\sigma_i^{P_i})_{P_i\in \mathcal{P},\,i\in N}$ OSP-implement $f$ with respect to $\mathcal{S}^\ast$. $\hfill \square $ 

\subsection{Two general results}

Proposition \ref{Elevator} establishes that for any social choice function $f$ the property of being OSP with respect to a given partition is inherited by all of its coarser partitions. 
Thus, in Example 1 above, $f$ is also OSP with respect to the coarser partition $\mathcal{S}=\{\{1,2,3\},\{4,5\}\}$ of $\mathcal{S}^{\ast}=\{\{1,2\},\{3\},\{4,5\}\}$.
We now state and prove Proposition \ref{Elevator}.

\begin{proposition}\label{Elevator} Let $f:\mathcal{D}\rightarrow A$ be OSP with respect to $\mathcal{S}^\ast$ and let $\mathcal{S}$ be a coarser partition of $\mathcal{S}^\ast$.
Then, $f:\mathcal{D}\rightarrow A$ is OSP with respect to $\mathcal{S}$.
\end{proposition}

\noindent \textbf{Proof. }Let $\Gamma $ and $(\sigma ^{R_i}_i)_{R_i\in \mathcal{D}_i\,,\,i\in N}$ be the extensive game form and the type-strategy profile that OSP-implement $f$ with respect to $\mathcal{S}^{\ast }$. 
Hence, they induce $f$. 
Thus, it only remains to be shown that $(\sigma ^{R_i}_i)_{R_i\in \mathcal{D}_i\,,\,i\in N}$ is obviously dominant with respect to $\mathcal{S}$ in $\Gamma $.

Fix $i\in N$ and $R_{i}\in \mathcal{D}_i$. 
To lighten the notation is this proof, we will write $\sigma _{i}$ instead of $\sigma _{i}^{R_{i}}$. 
Let $S\in \mathcal{S}$ and $S^{\ast}\in \mathcal{S}^{\ast }$ be such that $i\in S^{\ast }\subseteq S$. 
Fix a strategy $\sigma _{j}$ for all $j\in S\setminus \{i\}$ and let $\sigma _{i}^{\prime }\neq \sigma _{i}$.\smallskip

\noindent \textbf{Claim. }\;Let $I_{i}\in \mathcal{I}_{i}$ be such that $\sigma_{i}(I_{i})\neq \sigma _{i}^{\prime }(I_{i})$ and $\sigma_{i}(I_{i}^{\prime })=\sigma _{i}^{\prime }(I_{i}^{\prime })$ for all $I_{i}^{\prime }\prec I_{i}$.
Then,

\noindent (i) if $I_{i}(\sigma _{S},\sigma _{i}^{\prime })\in \alpha(\sigma_{S},\sigma _{i}^{\prime })$, then $I_{i}(\sigma_{S},\sigma_{i}^{\prime})\subseteq I_{i}(\sigma _{S^{\ast }},\sigma_{i}^{\prime })$, and

\noindent (ii) if $\overline{\sigma }_{-S}\in \Sigma _{-S}$, then $(\overline{\sigma }_{-S},\sigma _{S\setminus S^{\ast }})\in \Sigma _{-S^{\ast }}$.\smallskip

\noindent \textbf{Proof of the Claim. }To prove (i), let $I_{i}(\sigma_{S},\sigma _{i}^{\prime })\in \alpha (\sigma _{S},\sigma _{i}^{\prime })$ and $z_t\in I_{i}(\sigma _{S},\sigma _{i}^{\prime })$ be arbitrary. 
Then, the history $h=z_{0},\ldots ,z_{t}$ is compatible with $\sigma _{S}$. 
Hence, if $z_{t^{\prime }}\in Z_{j},$ with $t^{\prime }<t$ and $j\in S$, then $\sigma_{j}(z_{t^{\prime }})=z_{t^{\prime }+1}$. 
Therefore, as $S^{\ast }\subseteq S $, if $z_{t^{\prime }}\in Z_{j}$, with $t^{\prime }<t$ and $j\in S^{\ast }$, then $\sigma _{j}(z_{t^{\prime }})=z_{t^{\prime }+1}$. 
Therefore, $h=z_{0},\ldots ,z_{t}$ is compatible with $\sigma _{S^{\ast }}$. 
Hence $z_t\in I_{i}(\sigma _{S^{\ast }},\sigma _{i}^{\prime })$.

The proof of (ii) follows immediately from the observation that $S^{\ast}\subseteq S$. $\hfill \square $\smallskip

To proceed with the proof of Proposition \ref{Elevator}, let $I_{i}(\sigma_{S},\sigma _{i}^{\prime })\in \alpha (\sigma _{S},\sigma _{i}^{\prime })$ be given. 
By the claim above, 
\begin{eqnarray*}
\min_{R_{i}}\{x &\in &X\mid \exists \overline{\sigma}_{-S}\in \Sigma _{-S}\text{ and }z\in I_{i}(\sigma _{S},\sigma _{i}^{\prime })\text{ such that }x=g(z^{\Gamma }(z,(\sigma _{i},\sigma_{S\setminus \{i\}},\overline{\sigma }_{-S})))\} \\
R_{i}\min_{R_{i}}\{x &\in &X\mid \exists \overline{\sigma }_{-S^{\ast }}\in\Sigma _{-S^{\ast }}\text{ and }z\in I_{i}(\sigma _{S^{\ast }},\sigma_{i}^{\prime })\text{ such that }x=g(z^{\Gamma }(z,(\sigma _{i},\sigma_{S^{\ast }\setminus \{i\}},\overline{\sigma }_{-S^{\ast }})))\},
\end{eqnarray*}
because the first set of options, where the minimum is taken, is a subset of the second one, and 
\begin{eqnarray*}
\max_{R_{i}}\{x &\in &X\mid \exists \overline{\sigma }_{-S^{\ast }}\in\Sigma _{-S}^{\ast }\text{ and }z\in I_{i}(\sigma _{S^{\ast }},\sigma_{i}^{\prime })\text{ such that }x=g(z^{\Gamma }(z,(\sigma _{i}^{\prime},\sigma _{S^{\ast }\setminus\{i\}},\overline{\sigma }_{-S^{\ast }})))\} \\
R_{i}\max_{R_{i}}\{x &\in &X\mid \exists \overline{\sigma }_{-S}\in \Sigma_{-S}\text{ and }z\in I_{i}(\sigma _{S},\sigma _{i}^{\prime })\text{ such that }x=g(z^{\Gamma }(z,(\sigma _{i}^{\prime },\sigma _{S\setminus \{i\}},\overline{\sigma }_{-S})))\}
\end{eqnarray*}
because the first set of options, where the maximum is taken, contains the second one. 
Therefore, as $f$ is OSP with respect to $\mathcal{S}^{\ast }$, 
\begin{eqnarray*}
\min_{R_{i}}\{x &\in &X\mid \exists \overline{\sigma }_{-S^*}\in \Sigma_{-S^*} \text{ and }z\in I_{i}(\sigma _{S^*},\sigma_i^{\prime})\text{ such that }x=g(z^{\Gamma }(z,(\sigma _{i},\sigma _{S^*\setminus \{i\}},\overline{\sigma }_{-S^*})))\} \\
R_{i}\max_{R_{i}}\{x &\in &X\mid \exists \widehat{\sigma }_{-S^*}\in\Sigma_{-S^*}\text{ and }z\in I_{i}(\sigma_{S^*},\sigma_i^{\prime})\text{ such that }x=g(z^{\Gamma }(z,(\sigma _{i}^{\prime },\sigma _{S^*\setminus \{i\}},\widehat{\sigma }_{-S^*})))\}.
\end{eqnarray*}
Applying the transitivity of $R_i$, we obtain that 
\begin{eqnarray*}
\min_{R_{i}}\{x &\in &X\mid \exists \overline{\sigma }_{-S}\in \Sigma _{-S}\text{ and }z\in I_{i}(\sigma _{S},\sigma_i^{\prime})\text{ such that }x=g(z^{\Gamma }(z,(\sigma _{i},\sigma _{S\setminus\{i\}},\overline{\sigma }_{-S})))\} \\
R_{i}\max_{R_{i}}\{x &\in &X\mid \exists \widehat{\sigma }_{-S}\in\Sigma_{-S}\text{ and }z\in I_{i}(\sigma _{S},\sigma_i^{\prime})\text{ such that }x=g(z^{\Gamma }(z,(\sigma _{i}^{\prime },\sigma _{S\setminus\{i\}},\widehat{\sigma }_{-S})))\}.
\end{eqnarray*}
Thus, for all $x\in o(\sigma _S, \sigma ^{\prime}_i)$ and $y\in o^{\prime}(\sigma _S, \sigma ^{\prime}_i)$, 
\begin{equation*}
x\,R_i\,y.
\end{equation*}
Then, $\sigma _i^{R_i}$ is obviously dominant with respect to $\mathcal{S}$ in $\Gamma $ for $i$ with $R_i$.
Therefore, $f$ is OSP with respect to $\mathcal{S}$. \hfill $\blacksquare $\medskip

Given a partition $\mathcal{S}$ of the set of agents and a domain $\mathcal{D}=\mathcal{D}_1\times \cdots \times \mathcal{D}_n\subseteq \mathcal{R}^N$ of preferences, we define the class of finite extensive game forms $\mathcal{G}^{\mathcal{S}}$ through a finite sequence of steps.
Specifically, $\Gamma \in \mathcal{G}^\mathcal{S}$ if the following conditions hold.
\begin{itemize}
\item \underline{Step 1}: There exists $S^{1}\in \mathcal{S}$ such that agents in $S^{1}$ play only once and simultaneously, and for each $i\in S^1$, the set of available choices is a partition of $\mathcal{D}_i$.

\item \dots

Given $S^{1},\dots ,S^{k-1}$ identified in steps $1$ through $k-1$.

\item \underline{Step $k$}: For each non-terminal and commonly known history $h^{k-1}$ of Step $k-1$, there exists $S^{k}\in \mathcal{S}$ such that agents in $S^{k}$ play only once and simultaneously.
The set of available choices for each $i\in S^k$ is either a partition of $\mathcal{D}_i$, if $i$ has not played yet along $h^{k-1}$, or a partition of the subset of preferences chosen by $i$ last step $i$ has played along $h^{k-1}$, otherwise. 
Moreover, if agent $i\in S^k$ had only one available choice last step $k'<k$ where $i$ has played (which would imply that $S^{k'}=S^k$), then $i$ has the same singleton set of available choices in this Step $k$.
\end{itemize}

Observe that $S^k$ and $S^{k^{\prime}}$ may coincide for some pair of steps $k\neq k^{\prime }$. 
However, to be in $\mathcal{G}^{\mathcal{S}}$ the game $\Gamma$ has to finish after a finite number of steps. 

The game $\Gamma$ depicted in Figure 1 belongs to $\mathcal{G}^{\mathcal{S}^\ast}$ for $\mathcal{S}^\ast =\{\{1,2\},\{3\},\{4,5\}\}$, where $S^1=\{1,2\}$, $S^2=\{3\}$ for $h^1=(z_0,z_2,z_3)$ and $S^2=\{4,5\}$ for $\hat{h}^1=(z_0,z_1,z_4)$.

We say that $(\sigma^{R_i}_i)_{R_i\in \mathcal{D}_i}$ is a \textit{truth-telling type-strategy} of player $i$ in $\Gamma \in \mathcal{G}^{\mathcal{S}}$ if, for each $R_i\in\mathcal{D}_i$ and each information set $I_i\in \mathcal{I}_i$ such that there exits $a_i\in Ch(I_i)$ with $R_{i} \in a_i$, $\sigma _{i}^{R_{i}}(I_i)=a_i$; namely, $i$ always chooses the subset in the available partition of preferences that contains $R_i$, if such a subset exists.\footnote{Observe that this definition does not specify the choice of the strategy in an information set $I_i$ where there is no  $a_i\in Ch(I_i)$ containing $R_{i}$. 
In such information sets, the strategy can chose any available choice.}

\begin{theorem}
\label{Thm necessity} \label{SP-implements} 
Let $f:\mathcal{D}\rightarrow A$ be a social choice function and let $\mathcal{S}$ be a partition of $N$.
Suppose $\Gamma \in \mathcal{G}^{\mathcal{S}}$ and the truth-telling type-strategy profile $(\sigma _{i}^{R_{i}})_{R_i \in \mathcal{D}_i\,,\,i\in N}$ SP-implement $f$. 
Then, $\Gamma $ and $(\sigma _{i}^{R_{i}})_{R_i\in \mathcal{D}_i\,,\,i \in N}$ also OSP-implement $f$ with respect to $\mathcal{S}$.
\end{theorem}

\noindent \textbf{Proof. }Let $\Gamma \in \mathcal{G}^{\mathcal{S}}$ and $(\sigma _{i}^{R_{i}})_{R_{i}\in \mathcal{D}_{i}\,,\,i\in N}$ be the game and the truth-telling type-strategy profile that SP-implement $f$. 
Hence, for each $R\in \mathcal{D}$, (i) $f(R)=g(z^{\Gamma }(z_{0},\sigma ^{R}))$ and (ii) for all $i\in N$, $\sigma_{i}^{R_{i}}$ is weakly dominant in $\Gamma $ for $i$ with $R_{i}$.
Fix $i\in N$ and let $S\in \mathcal{S}$ be such that $i\in S$.
Let $\sigma ^{\prime }_i\in \Sigma _i\setminus \{\sigma_i^{R_i}\}$ be any deviating strategy of agent $i$.
Fix an strategy, $\sigma_{S\setminus\{i\}}$, for agents in $S\setminus\{i\}$, let $\sigma_{S}=(\sigma_{S\setminus\{i\}},\sigma_i^{R_i})$, and let $I_i(\sigma_{S},\sigma'_i)\in\alpha(\sigma_{S},\sigma'_i)$. 

Select any $\theta _{-S},\theta _{-S}^{\prime }\in \Sigma_{-S}$, $z,z' \in I_i(\sigma_{S},\sigma'_i)$ and $y,y^{\prime}\in A$ for which 
\begin{equation*}
x\,R_i\,y=g(z^{\Gamma }(z,(\sigma _i^{R_i}, \sigma_{S\setminus \{i\}},\theta _{-S}))),
\end{equation*}
for all $x\in o(\sigma_{S},\sigma ^{\prime }_i)$ and 
\begin{equation*}
y^{\prime}=g(z^{\Gamma }(z',(\sigma ^{\prime}_{i},\sigma _{S\setminus \{i\}},\theta ^{\prime} _{-S})))\,R_i\,x^{\prime },
\end{equation*}
for all $x^{\prime }\in o^{\prime}(\sigma_{S},\sigma ^{\prime }_i)$.

Namely, given $\sigma_{S}$ and $\sigma ^{\prime }_i$, $\theta_{-S}$ and $\theta ^{\prime}_{-S}$ are two profiles of strategies of the agents not in $S$ that induce, respectively, alternatives $y$ and $y^{\prime}$.
These alternatives are among the least or most preferred alternatives, respectively, in the sets of options left by $\sigma_{S}$ together with $\sigma^{\prime }_i$ at the earliest point of departure $I_i(\sigma_{S},\sigma'_i)$.
Without loss of generality, by definition of information sets in the game, we can modify  $\theta_{-S}$ and $\theta ^{\prime}_{-S}$ to ensure that $z$ and $z'$ are compatible with $\theta_{-S}$ and $\theta ^{\prime}_{-S}$, respectively. 
Then, we can assume that  
\begin{equation*}
y=g(z^{\Gamma }(z_0,(\sigma _i^{R_i}, \sigma_{S\setminus \{i\}},\theta _{-S}))),
\end{equation*}
and 
\begin{equation*}
y^{\prime}=g(z^{\Gamma }(z_0,(\sigma ^{\prime}_{i},\sigma _{S\setminus \{i\}},\theta ^{\prime} _{-S}))).
\end{equation*}

Define, for each $j\notin S$, the behavioral strategy $\widehat{\sigma}_j$ such that, for each $z\in Z_j$, 
\begin{equation*}
\widehat{\sigma }_{j}(z)=\left\{ 
\begin{tabular}{cc}
$\theta _{j}$ & if agents in $S$ play in the history towards $z$ according to $(\sigma ^{R_i}_{i},\sigma ^{R_{{S\setminus \{i\}}}}_{S\setminus \{i\}})$\medskip \\ 
\multicolumn{1}{l}{$\theta _{j}^{\prime }$} & \multicolumn{1}{l}{if agents in $S$ play in the history towards $z$ according to $(\sigma_{i}^{\prime },\sigma ^{R_{{S\setminus \{i\}}}}_{S\setminus \{i\}})$.}
\end{tabular}
\right.
\end{equation*}

\noindent Then, for all $x\in o(\sigma_S^{R_{S}},\sigma ^{\prime }_i)$ and $x^{\prime}\in o^{\prime}(\sigma_S^{R_{S}},\sigma ^{\prime }_i)$, 
\begin{equation*}
\begin{tabular}{cll}
$x$\hspace{-0.20in} & $R_i\,y=g(z^{\Gamma }(z_0,(\sigma _i^{R_i}, \sigma_{S\setminus \{i\}},\theta _{-S})))$ & by definitions of $\theta _{-S}$ and $y$\smallskip \\ 
& $=g(z^{\Gamma }(z_0,(\sigma _i^{R_i}, \sigma_{S\setminus \{i\}},\widehat{\sigma }_{-S})))$ & by definition of $\widehat{\sigma }_{-S}$ \\ 
& $R_{i}$ &  \\ 
& $g(z^{\Gamma }(z_0,(\sigma _{i}^{\prime },\sigma _{S\setminus \{i\}}, \widehat{\sigma }_{-S})))$ & because $\sigma ^{R_i}_i$ is a dominant strategy in $\Gamma$\smallskip \\ 
& $=g(z^{\Gamma }(z_0,(\sigma _{i}^{\prime },\sigma _{S\setminus \{i\}},\theta _{-S}^{\prime })))$ & by definition of $\widehat{\sigma }_{-S}$\smallskip \\ 
& $=y^{\prime }\,R_i\,x^{\prime}$ & by definitions of $\theta ^{\prime}_{-S}$ and $y^{\prime}$.
\end{tabular}%
\end{equation*}
Therefore, $\sigma _{i}^{R_{i}}$ is obviously dominant with respect to $\mathcal{S}$ in $\Gamma $ for $i$ with $R_{i}$ and $\Gamma $ and $(\sigma _{i}^{R_{i}})_{R_i\in \mathcal{D}_i\,,\,i \in N}$ also OSP-implement $f$ with respect to $\mathcal{S}$.
$\hfill \blacksquare $

\subsection{Round table mechanisms}

Before concluding this section, we note that, similar to the OSP-implementation in our framework, OSP-implementing a social choice function with respect to a partition can be achieved by focusing on round table mechanisms, with or without perfect information (see Mackenzie (2020) for the case of perfect information).
This approach stems from two key ideas, which extend the arguments for OSP-implementation to OSP-implementation with respect to a partition.

The first idea relates to the pruning principle (see, for instance, Li (2017) and Ashlagi and Gonczarowski (2018)). 
Suppose the pair $(\Gamma ,(\sigma _{i}^{R_{i}})_{R_i \in \mathcal{D}_i\,,\,i\in N})$, consisting of an extensive game form and a type-strategy profile, OSP-implements the social choice function $f:\mathcal{D}\rightarrow A$.
Delete the last parts of the paths in $\Gamma $ that are never played when agents use $(\sigma _{i}^{R_{i}})_{R_i \in \mathcal{D}_i\,,\,i\in N}$, denoting this pruned game by $\widehat{\Gamma }$ and the restriction of $(\sigma _{i}^{R_{i}})_{R_i \in \mathcal{D}_i\,,\,i\in N}$ to $\widehat{\Gamma }$ by $(\widehat{\sigma } _{i}^{R_{i}})_{R_i \in \mathcal{D}_i\,,\,i\in N}$. 
It is evident that the pair $(\widehat{\Gamma},(\widehat{\sigma } _{i}^{R_{i}})_{R_i \in \mathcal{D}_i\,,\,i\in N})$ also OSP-implements $f$, as after pruning $\Gamma $, the worst-case from continuing can only improve, and the best-case from deviating can only worsen.

The second idea involves the relabeling of the choices in $\widehat{\Gamma }$, as proposed by Mackenzie (2020).
For each agent $i$, at every history where $i$ must play, and for each available choice to $i$, relabel that choice with the collection of preferences $\widehat {R}_i\in \mathcal{D}_i$ whose corresponding part of $i$'s type-strategy $(\sigma_i^{R_i})_{R_i\in \mathcal{D}_i\,,\,i\in N}$ is compatible with the history. Additionally, the corresponding strategies $\sigma_i^{\widehat{R}_i}$ select that choice.

The extensive game form obtained after pruning and relabeling of choices is termed a \emph{round table mechanism}. It remains an extensive game form, potentially with imperfect information due to the role of the partition. In this form, the sets of choices are non-empty subsets of preferences that satisfy the following properties: (a) the choices at any information set are disjoint subsets of preferences, (b) when a player $i$ plays for the first time, the set of choices is a partition of $\mathcal{D}_i$, and (c) subsequently, at an information set $I_i$, the union of available choices is the intersection of the choices made by agent $i$ at all predecessor information sets leading to $I_i$.

Note that the extensive game forms utilized in Theorem \ref{SP-implements} and the application to extended majority voting rules with two alternatives (as demonstrated in Theorems \ref{anonymity} and \ref{SA} in Section 4) may all be considered round table mechanisms with imperfect information. In the proofs of the latter two theorems, we will employ the following remark, which is applicable to any scenario where, for all $i\in N$, $\lvert \mathcal{D}_i\rvert=2$; for instance, in cases where there are two alternatives, denoted as $x$ and $y$, and agents' preferences are strict.

\begin{remark}\label{RTM}
Let $\Gamma $ be an extensive game form that OSP-implements the social choice function $f:\mathcal{D}\rightarrow \{x,y\}$ with respect to a partition $\mathcal{S}$, where, for all $i\in N$, $\mathcal{D}_i =\{P^x,P^y\}$, $x\,P^x\,y$ and $y\,P^y\,x$. 
Assume that no agent is dummy in $f$.
Then, we can assume that $\Gamma $ is a round table mechanism.
Moreover, along any complete history $h$, each agent $i$ plays at most once, choosing one preference from the choice set $\{\{P^x\},\{P^y\}\}$, and $h$ can be written as $h=z_0,\dots ,z_{\ell^\prime },\dots, z_\ell$, where $\ell \leq n$ and $z_\ell\in Z_T$ is a terminal node.
\hfill $\square$ 
\end{remark}

We now argue why Remark \ref{RTM} holds. Along any complete history $h=(z_0,\dots, z_T)\in H$ where agent $i$ plays, the sequence of choice sets for agent $i$, denoted by $\{Ch(z_t)\}_{t\in \Theta_i}$, where $\Theta_i=\{t\in\{0,\dots,T\}\mid z_t\in Z_i\}$, begins with a subsequence of singleton choice sets $\{\{P^{x},\,P^{y}\}\}$. If such a subsequence exists, there exists $\widehat{t}\in \Theta_i$ such that $Ch(z_{\widehat{t}})=\{\{P^{x}\},\{P^{y}\}\}$, and it continues with either the subsequence of singleton choice sets $\{\{P^{x}\}\}$ if the node $z_{\widehat{t}+1}$ is associated with the choice $a_i^{\widehat{t}}=\{P^x\}$ at $z_{\widehat{t}}$, or with the subsequence of singleton choice sets $\{\{P^{y}\}\}$ if the node $z_{\widehat{t}+1}$ is associated with the choice $a_i^{\widehat{t}}=\{P^y\}$ at $z_{\widehat{t}}$.
Since $\Gamma$ induces $f$, the initial and final subsequences of singleton sets can be removed, leaving only the node $z_{\widehat{t}}$ in the history $h$ whose choice set is $\{\{P^{x}\},\{P^{y}\}\}$.
Therefore, without loss of generality, we can assume that along any complete history $h\in H$, each agent $i$ plays at most once (i.e., for each $i$, $\Theta_i$ is either empty or a singleton set), choosing a preference from the set $\{\{P^{x}\},\{P^{y}\}\}$.
Correspondingly, $h$ can be represented as $h=z_0,\dots ,z_{\ell^\prime },\dots, z_\ell$, where $\ell \leq n$ and $z_\ell\in Z_T$ is a terminal node.
For any $\ell^\prime <\ell$, we will refer to the agent who owns node $z_{\ell^\prime }$ as $\mathcal{N}(z_{\ell^\prime })$.

In general, the extensive game form $\Gamma$ that OSP-implements the social choice function $f$ with respect to a partition $\mathcal{S}$ must have imperfect information.
To see why, consider the following argument.
By definition, if $\Gamma$ were to OSP-implement $f$ with perfect information, then it would also SP-implement $f$.
However, Mackenzie (2020) establishes that SP-implementation with perfect information is equivalent to OSP-implementation. Since OSP-implementation with respect to a partition is strictly stronger than OSP-implementation alone, $\Gamma$ cannot have perfect information.
For example, in the application in Subsection 4.2 with two alternatives, there are instances of anonymous social choice functions that are OSP-implementable with respect to $\mathcal{S}$ but, according to Arribillaga, Massó, and Neme (2020), they are not OSP-implementable.
This indicates that a certain level of imperfect information is required for OSP-implementation with respect to $\mathcal{S}$.

\section{Obviously strategy-proof anonymous extended majority voting rules with respect to a partition}

We apply the notion of OSP with respect to a partition to the simplest social choice problem where there are only two alternatives and agents' preferences are strict.

In the first subsection, we define the class of Extended Majority Voting Rules (EMVRs).
This class coincides with the family of all strategy-proof social choice functions within this setting.
Thus, our search for social choice functions that are OSP with respect to a partition is confined to this class. 
In the second subsection, we provide a characterization of the family of  anonymous extended majority voting rules that are obviously strategy-proof with respect to a partition. Additionally, these rules satisfy two alternative notions of anonymity.\footnote{In Arribillaga, Massó and Neme (2024a) we identify a subclass of extended majority voting rules, which may not be anonymous, that are OSP-implementable with respect to a partition.
The extensive game forms used for their OSP-implementation possess a particular  simple feature.\smallskip}\medskip

\subsection{Extended majority voting rules: Preliminaries}

Let $A=\{x,y\}$ be the set of alternatives and $\mathcal{P}$ be the set of the two strict preferences over $A$; namely, $\mathcal{P}=\{P^x,P^y\}$, where $x\,P^x\,y$ and $y\,P^y\,x$.
A social choice function is then a mapping $f:\{P^x,P^y\}^N\rightarrow \{x,y\}$.

To characterize the class of all obviously strategy-proof social choice functions with respect to a partition, we can simplify the process by focusing on the subset of strategy-proof social choice functions, which can be described using the notion of a committee.

Let $2^{N}$ denote the family of all subsets of $N$ (called coalitions). 
A family $\mathcal{C}\subset 2^{N}$ of coalitions is a \emph{committee} if it is (coalition) monotonic, meaning that for each pair $T,T^{\prime}\subseteq N$ such that $T\in \mathcal{C}$ and $T\subsetneq T^{\prime}$, we have $T^{\prime}\in \mathcal{C}$.
Coalitions in $\mathcal{C}$ are called \emph{winning}. 
Given $\mathcal{C}$, denote by $\mathcal{C}_{m}$ the family of \emph{minimal winning coalitions} of $\mathcal{C}$, defined as 
\begin{equation*}
\mathcal{C}_{m}=\{T\in \mathcal{C}\mid \text{there is no }T^{\prime }\in \mathcal{C}\text{ such that }T^{\prime }\subsetneq T\}.
\end{equation*}
Observe that, due to the monotonicity property of a committee, specifying $\mathcal{C}_{m}$ is sufficient to completely determine $\mathcal{C}$.\medskip

\begin{definition}\label{DefEMVR}
A social choice function $f:\{P^x,P^y\}^N\rightarrow \{x,y\}$ is an \emph{extended majority voting rule (EMVR)} if there exists a committee $\mathcal{C}^x$ with the property that, for all $P\in \{P^x,P^y\}^N$, 
\begin{equation}
f(P)=x\text{ if and only if }\{i\in N\mid P_{i}=P^x\}\in \mathcal{C}^x.
\label{GMV}
\end{equation}
\end{definition}

We denote the extended majority voting rule whose associated committee is $\mathcal{C}^x$ by $f_{\mathcal{C}^x}$.
Before proceeding, two remarks about the definition of an EMVR are in order.

First, the above definition is relative to a committee for $x$ (this is reflected in the use of the notation $\mathcal{C}^x$). 
It is possible to define the symmetric condition of (\ref{GMV}) relative to a committee for $y$, denoted by $\mathcal{C}^y$, by replacing $x$ with $y$ everywhere in (\ref{GMV}). 
Then, it is easy to show that $\mathcal{C}^x$ and $\mathcal{C}^y$ define the same $f$ if and only if 
\begin{equation}
T\in \mathcal{C}^{y}\text{ if and only if }T\cap T^{\prime }\neq \emptyset \text{ for all }T^{\prime }\in \mathcal{C}^{x}.
\label{Equivalence of committees}
\end{equation}
This condition ensures that the committees for $x$ and $y$ are equivalent in defining the same social choice function.
Additionally, we say that an agent $i$ is a \emph{dummy} in $\mathcal{C}$ if there does not exist $M\in \mathcal{C}_m$ such that $i\in M$; otherwise, $i$ is \emph{non-dummy} in $\mathcal{C}$.

Second, if the EMVR is onto then its associated committee $\mathcal{C}$ is not trivial (\emph{i.e.}, $\emptyset \notin \mathcal{C}\neq \emptyset $).
However, if the EMVR is not onto and therefore is constant, then $\emptyset\in \mathcal{C}$ if it always elects $x$, and $\mathcal{C}=\emptyset $ if it always elects $y $. 
Since constant social choice functions are obviously strategy-proof with respect to any partition, we will assume that all committees under consideration are not trivial, and accordingly, their associated EMVRs are onto.

We state as a remark the characterization of the class of all EMVRs in this simple context (this follows from a more general result in Barberà, Sonnenschein and Zhou (1991)).

\begin{remark}\label{RemSP-EMVR}
    A social choice function $f:\{P^x,P^y\}^N \rightarrow \{x,y\}$ is strategy-proof if and only if $f$ is an \emph{EMVR}; namely, there exists a committee $\mathcal{C}^x$ such that $f=f_{\mathcal{C}^x}$.\hfill $\square$ 
\end{remark}

We are interested in the social choice functions $f$ that are OSP with respect to a partition and anonymous.
However, in light of Remark 3, we can focus on the committees associated with $f$, along with their respective anonymity properties.

\subsection{Anonymity}

We characterize the class of committees $\mathcal{C}^x$ for which the corresponding EMVRs $f_{\mathcal{C}^{x}}:\{P^x,P^y\}^N\rightarrow \{x,y\}$ are OSP with respect to a partition and additionally satisfy two alternative notions of anonymity. 
Theorem \ref{anonymity} addresses anonymity relative to a partition (where the chosen alternative does not change after agents' names are permuted only among the members belonging to the same subset of the partition), while Theorem \ref{SA} deals with strong anonymity (where the chosen alternative does not change after agents' names are permuted in any way).

\subsubsection{Anonymity relative to a partition}

Let $\mathcal{S}$ be a partition of $N$, and let $\Pi ^{\mathcal{S}}$ be the set of all bijections $\pi ^{\mathcal{S}}:N\rightarrow N$ that only swap agents within each element of $\mathcal{S}$; namely, $\pi ^{\mathcal{S}}\in \Pi ^{\mathcal{S}}$ if and only if, for each $S\in \mathcal{S}$, $\pi ^{\mathcal{S}}(S)=S$.

A committee $\mathcal{C}^{x}$ is \emph{anonymous relative to a partition} $\mathcal{S}$ if (i) it does not have dummy agents and (ii) for all $\pi ^{\mathcal{S}}\in \Pi ^{\mathcal{S}}$ and $M\in \mathcal{C}^{x}$, $\pi ^{\mathcal{S}}(M)\in \mathcal{C}^{x}$.\footnote{A committee satisfying only condition (ii) in this definition could have dummy agents (for example, when all agents in $S$ are dummy for some $S\in \mathcal{S}$) and condition (i) explicitly excludes this possibility. 
Of course, attributing any property of anonymity to a committee with both dummy and non-dummy agents would sound strange.\smallskip}

Our goal is to characterize the class of committees that are anonymous relative to a partition $\mathcal{S}$ and whose associated EMVRs are OSP with respect to $\mathcal{S}$.
We will accomplish this by employing a straightforward class of extensive game forms, which we describe shortly after introducing some additional notation.

First, to denote a specific order of elements within a partition $\mathcal{S}$ comprising $K$ subsets of $N$, we use $\mathcal{S}^o=(S_1,\dots,S_K)$ and refer to it as an \emph{ordered partition}.

Second, given an ordered partition $\mathcal{S}^o=(S_{1},\dots,S_{K})$, we say that a vector $Q=(q_{1},\dots ,q_{K})\in \{0,1,\dots,n\}^K$ is a \emph{compatible} vector of quotas (with respect to $\mathcal{S}^o$) if, for all $k\in \{1,\dots, K-1\}$, $q_k\leq \lvert S_k \rvert$ and $q_K<\left\vert S_{K}\right\vert$.

Given $\mathcal{S}^{o}$ and a compatible vector of quotas $Q$, we define an extensive game form $\Gamma _{\mathcal{S}^{o},\,Q}$ in $\mathcal{F}^{\mathcal{S}^{o},\,Q} $, which is a subclass of $\mathcal{G}^{\mathcal{S}}$ containing all extensive game forms obtained through a $[\mathcal{S}^{o},Q]$-process, defined below.

\begin{center}    \underline{$[\mathcal{S}^{o},Q]$-process}
\end{center}

\begin{itemize}

\item \underline{Step $k$} ($1\leq k<K$): 
For each non-terminal and commonly known history $h^{k-1}$ at the end of Step $k-1$, agents in $S_{k}$ play only once and simultaneously.\footnote{Recall that $h^0$ is the initial and empty history.
By convention, Step 0 is empty.\smallskip} 
The set of available choices $Ch(I_i)$ for each $i\in S_k$ at each $I_i\in\mathcal{I}_i$ is equal to the partition $\{\{P^x\},\{P^y\}\}$. 
Let $h^{k}$ be a given history at the end of Step $k$. 
Then, (i) $h^{k}$ is terminal and the outcome of $\Gamma _{\mathcal{S}^{o},\,Q}$ is $x$ if strictly more than $q_k$ agents in $S_k$ have chosen $\{P^x\}$ along $h^k$, (ii) $h^{k}$ is terminal and the outcome of $\Gamma _{\mathcal{S}^{o},\,Q}$ is $y$ if strictly fewer than $q_k$ agents in $S_k$ have chosen $\{P^x\}$ along $h^k$, and (iii) $h^{k}$ is non-terminal if exactly $q_k$ agents in $S_k$ have chosen $\{P^x\}$ along $h^k$, in which case proceed to Step $k+1$.

\item \underline{Step $K$}: For each non-terminal and commonly known history $h^{K-1}$ at the end of Step $K-1$, agents in $S_{K}$ play only once and simultaneously. 
The set of available choices $Ch(I_i)$ for each $i\in S_{K}$ at each $I_i\in\mathcal{I}_i$ is equal to the partition $\{\{P^{x}\},\{P^{y}\}\}$. 
Let $h^{K}$ be a given history at the end of Step $K$. 
Then, (i) $h^{K}$ is terminal and the outcome of $\Gamma _{\mathcal{S}^{o},\,Q}$ is $x$ if strictly more than $q_{K}$ agents in $S_{K}$ have chosen $\{P^{x}\}$ along $h^{K}$ and (ii) $h^{K}$ is terminal and the outcome of $\Gamma _{\mathcal{S}^{o},\,Q}$ is $y$ if fewer than or exactly $q_{K}$ agents in $S_{K}$ have chosen $\{P^{x}\}$ along $h^{K}$.
\end{itemize}

Figure 2 depicts Step 1 of the extensive game form $\Gamma _{\mathcal{S}^{o},\,Q}$, where $N^x(h^1)$ denotes the number of agents that chose $P^x$ along the history $h^1$, the outcome of Step 1. 

\begin{center}
\begin{tikzpicture}[line cap=round,line join=round,>=triangle 45,x=0.8cm,y=.6cm]
\draw (-0.3,1) node[anchor=north west] {$i$};
\draw (-0.6,-0.2) node[anchor=north west] {$z_0$};
\draw (2.5,6.2) node[anchor=north west] {$S_1$};
\draw (6.3,7.2) node[anchor=north west] {$x$};
\draw (6.3,6.2) node[anchor=north west] {$x$};
\draw (6.3,5.2) node[anchor=north west] {$x$};
\draw (6.3,3.2) node[anchor=north west] {$x$};
\draw (6.3,4.2) node[anchor=north west] {$x$};
\draw (6.3,-4.8) node[anchor=north west] {$y$};
\draw (6.3,-3.8) node[anchor=north west] {$y$};
\draw (6.3,-5.8) node[anchor=north west] {$y$};
\draw (6.3,-6.8) node[anchor=north west] {$y$};
\draw (6.3,0.2) node[anchor=north west] {$S_2$};
\draw [color=red](7.4,-4.8) node[anchor=north west] {$N^x(h^1)<q_1$};
\draw [color=blue](7.4,5.2) node[anchor=north west] {$N^x(h^1)>q_1$};
\draw (7.4,0.2) node[anchor=north west] {$N^x(h^1)=q_1$};
\draw [line width=0.8pt,dash pattern=on 1pt off 2pt on 4pt off 4pt,color=blue] (0.,0.)-- (6.,7.);
\draw [line width=0.8pt,dash pattern=on 1pt off 2pt on 4pt off 4pt,color=blue] (0.,0.)-- (6.,3.);
\draw [line width=0.8pt,dash pattern=on 1pt off 2pt on 4pt off 4pt,color=red] (0.,0.)-- (6.,-4.);
\draw [line width=0.8pt,dash pattern=on 1pt off 2pt on 4pt off 4pt,color=red] (0.,0.)-- (6.,-7.);
\draw [line width=0.8pt,dash pattern=on 1pt off 2pt on 4pt off 4pt] (0.,0.)-- (6.,1.);
\draw [line width=0.8pt,dash pattern=on 1pt off 2pt on 4pt off 4pt] (0.,0.)-- (6.,-1.);
\draw [line width=0.8pt,color=blue] (0.,0.)-- (2.,1.3);
\draw [line width=0.8pt,color=blue] (2.,1.3)-- (3.,2.5);
\draw [line width=0.8pt,color=blue] (4.,2.7)-- (5.,3.7);
\draw [line width=0.8pt,color=blue] (5.,3.7)-- (6.,4.);
\draw [line width=0.8pt,dotted,color=blue] (3.,2.5)-- (4.,2.7);
\draw [line width=0.8pt] (0.,0.)-- (2.,0.1);
\draw [line width=0.8pt] (3.,-0.4)-- (2.,0.1);
\draw [line width=0.8pt] (4.,-0.2)-- (5.,0.4);
\draw [line width=0.8pt] (5.,0.4)-- (6.,0.);
\draw [line width=0.8pt,dotted] (3.,-0.4)-- (4.,-0.2);
\draw [line width=0.8pt,dotted,color=red] (3.,-2.2)-- (4.,-4.);
\draw [line width=0.8pt,color=red] (0.,0.)-- (2.,-2.);
\draw [line width=0.8pt,color=red] (2.,-2.)-- (3.,-2.2);
\draw [line width=0.8pt,color=red] (4.,-4.)-- (5.,-4.2);
\draw [line width=0.8pt,color=red] (5.,-4.2)-- (6.,-5.);
\begin{scriptsize}
\draw [fill=blue] (0.,0.) circle (1.0pt);
\draw [fill=blue] (6.,7.) circle (1.0pt);
\draw [fill=blue] (6.,-7.) circle (1.0pt);
\draw [fill=blue] (6.,3.) circle (1.0pt);
\draw [fill=blue] (2.,1.3) circle (1.0pt);
\draw [fill=blue] (3.,2.5) circle (1.0pt);
\draw [fill=blue] (4.,2.7) circle (1.0pt);
\draw [fill=blue] (5.,3.7) circle (1.0pt);
\draw [fill=blue] (6.,4.) circle (1.0pt);
\draw [fill=blue] (2.,-2.) circle (1.0pt);
\draw [fill=blue] (3.,-2.2) circle (1.0pt);
\draw [fill=blue] (4.,-4.) circle (1.0pt);
\draw [fill=blue] (5.,-4.2) circle (1.0pt);
\draw [fill=blue] (6.,1.) circle (1.0pt);
\draw [fill=blue] (6.,-1.) circle (1.0pt);
\draw [fill=blue] (6.,-0.5) circle (1.0pt);
\draw [fill=blue] (2.,0.1) circle (1.0pt);
\draw [fill=blue] (3.,-0.4) circle (1.0pt);
\draw [fill=blue] (5.,0.4) circle (1.0pt);
\draw [fill=blue] (6.,0.) circle (1.0pt);
\draw [fill=blue] (6.,-4.) circle (1.0pt);
\draw [fill=blue] (6.,5.) circle (1.0pt);
\draw [fill=blue] (6.,6.) circle (1.0pt);
\draw [fill=blue] (6.,0.5) circle (1.0pt);
\draw [fill=blue] (4.,-0.2) circle (1.0pt);
\draw [fill=blue] (6.,-4.) circle (1.0pt);
\draw [fill=blue] (6.,-5.) circle (1.0pt);
\draw [fill=blue] (6.,-4.) circle (1.0pt);
\draw [fill=blue] (6.,-6.) circle (1.0pt);
\end{scriptsize}
\end{tikzpicture}

Figure 2: Step 1 of the extensive game form $\Gamma _{\mathcal{S}^{o},\,Q}$\medskip
\end{center}

\noindent The set of histories induced by all plays of agents in $S_1$ can be partitioned into three subsets.
Firstly, there are histories $h^1$ where the number of agents that chose $x$ along $h^1$ is strictly greater than the quota $q_1$ (\textit{i.e.} $N^x(h^1)>q_1$).
These histories, depicted in blue, are all terminal, and their associated outcome is $x$.
Secondly, there are histories $h^1$ where the number of agents that chose $x$ along $h^1$ is strictly less than the quota $q_1$ (\textit{i.e.} $N^x(h^1)<q_1$).
These histories, depicted in red, are all terminal and their associated outcome is $y$.
Thirdly, there are histories $h^1$ where the number of agents that chose $x$ along $h^1$ is equal to the quota $q_1$ (\textit{i.e.} $N^x(h^1)=q_1$). 
These histories, depicted in black, are all non-terminal and agents in $S_2$ proceed to play in Step 2, given $h^1$.

For each non-terminal history outcome of Step 1, Step 2 begins with agents in $S_2$ simultaneously making their choices. 
Similarly, for each non-terminal history, outcome of Step $k-1$, with $1 < k < K$, Step $k$ begins with agents in $S_k$ simultaneously making their choices.
Each subsequent step replicates the structure of Step 1. 
Step $K$ is similar to the previous steps, but with the difference that all histories where exactly $q_{K}$ from $S_{K}$ chose $\{P^{x}\}$ are now terminal and result in the outcome $y$.

Let $\Gamma _{\mathcal{S}^{o},\,Q}\in \mathcal{F}^{\mathcal{S}^{o},\,Q}$ be an extensive game form obtained through a $[\mathcal{S}^{o},\,Q]$-process.\footnote{Note that $\mathcal{F}^{\mathcal{S}^{o},\,Q} $ contains several extensive game forms, but this multiplicity is inconsequential as it arises from different ordering in which agents within the same element of the partition play.\smallskip}

Consider the truth-telling type-strategy $(\sigma _{i}^{P_{i}})_{P_i \in \mathcal{P}\,,\,i\in N}$, where for every $i\in N$ and every $I_{i}\in \mathcal{I}_{i}$, $\sigma _{i}^{P_{i}}(I_{i})=\{P^{x}\}$ if $P_{i}=P^{x}$ and $\sigma_{i}^{P_{i}}(I_{i})=\{P^{y}\}$ if $P_{i}=P^{y}$. 

In the proof of Theorem \ref{anonymity} we will use Lemma \ref{OD} which states that for each agent truth-telling is an obviously dominant strategy with respect to $\mathcal{S}$ in $\Gamma _{\mathcal{S}^{o},\,Q}$. 

\begin{lemma}\label{OD}
Let $\Gamma _{\mathcal{S}^{o},\,Q}\in \mathcal{F}^{\mathcal{S}^{o},\,Q}$ be an extensive game form obtained through a $[\mathcal{S}^{o},\,Q]$-process.
Then, the truth-telling type-strategy $(\sigma _{i}^{P_{i}})_{P_i \in \mathcal{P}\,,\,i\in N}$ is obviously dominant with respect to $\mathcal{S}$.
  
\end{lemma}

\noindent \textbf{Proof.}
Consider an arbitrary agent $i$ playing at the unique information set $I_i$, with $S_k$ being the set of agents playing simultaneously with $i$ at step $k$ of $\Gamma _{\mathcal{S}^{o},\,Q}$.
Since agents in $S_k$ are called to play in $\Gamma _{\mathcal{S}^{o},\,Q}$, for each $t\in\{1,\dots,k-1\}$, the number of agents in $S_t$ choosing $\{P^x\}$ is equal to $q_t$. 
Assume first that $i$`s preference is $P^x$. 
The truth-telling strategy for agent $i$ is to choose $\{P^x\}$. 
Fix an arbitrary strategy for the agents in $S_k$, where $i$ truth-tells and chooses $\{P^x\}$.
If the outcome of $\Gamma _{\mathcal{S}^{o},\,Q}$ is $x$, the worst outcome of truth-telling is $x$. 
If agent $i$ deviates and chooses $\{P^y\}$, the outcome can either be $x$ or $y$.
In either case, the best outcome of deviating is not better than $x$.
If the outcome of $\Gamma _{\mathcal{S}^{o},\,Q}$ is $y$, the worst outcome of truth-telling is $y$.
If agent $i$ deviates and chooses $\{P^y\}$, the outcome remains $y$, and the best outcome of deviating is not better than the worst outcome of truth-telling, which is $y$. 
If the game moves to step $k+1$, the truth-telling strategy induces either $x$ or $y$, with the worst outcome being $y$.
If agent $i$ deviates and chooses $\{P^y\}$, the game $\Gamma _{\mathcal{S}^{o},\,Q}$ does not move to step $k+1$ and the outcome is $y$, which is the best outcome of deviating.

In all cases, the worst preferred outcome of truth-telling is at least as good as the most preferred outcome of deviating.
Similarly, if $i$'s preference is $P^y$. 
Thus, in both cases, the truth-telling is indeed an obviously dominant strategy with respect to $\mathcal{S}$ in $\Gamma_{\mathcal{S}^o,\,Q}$.
\hfill $\blacksquare $\medskip

Before presenting the key necessary and sufficient condition identified in Theorem \ref{anonymity}, we present an example of an extensive game form $\Gamma _{\mathcal{S}^{o},\,Q}\in \mathcal{F}^{\mathcal{S}^{o},\,Q}$ that is obtained through a $[\mathcal{S}^{o},\,Q]$-process and identify the EMVR induced by this game and its corresponding truth-telling type-strategy profile.\medskip

\noindent \textbf{Example 2. }Let $N=\{1,\dots,10\}$ be the set of agents, $\mathcal{S}^o=(\{1,2,3\},\{4,5,6,7,8\},\{9,10\})$ be an ordered partition of $N$ and $Q=(q_1,q_2,q_3)=(2,5,0)$ be a compatible vector of quotas. 
The game $\Gamma _{\mathcal{S}^{o},\,Q}$ is obtained through the following $[\mathcal{S}^o,\,Q]$-process.
\begin{itemize}
    
\item \underline{Step $1$}: Agents in $S_1=\{1,2,3\}$ start by simultaneously choosing one in $\{\{P^x\},\{P^y\}\}$, referred to as voting for $x$ or for $y$, respectively. 
If all three agents vote for $x$ (the number of votes for $x$ is strictly greater than the quota $q_1$), the outcome is $x$.
If the number of votes for $x$ is either 0 or 1 (strictly less than the quota $q_1$), the outcome is $y$.
If exactly 2 two agents vote for $x$ (the quota $q_1$), the game moves to Step 2, where the outcome depends on the votes of agents in $S_2$, and eventually in $S_3$.

\item \underline{Step $2$}: For each non-terminal history $h^1$ at the end of Step 1, agents in $S_2$ simultaneously vote for $x$ or for $y$.
If the number of votes for $x$ is strictly less than 5 (the quota $q_2$), the outcome is $y$.
If all five agents in $S_2$ vote for $x$ (the quota $q_2$), the game proceeds to Step $3$, where the outcome depends on the votes of agents in $S_3$.

\item \underline{Step $3$}: For each non-terminal history $h^2$ at the end of Step 2, agents in $S_3$ simultaneously vote for $x$ or for $y$.
If either 1 or 2 agents vote for $x$ (numbers strictly greater than the quota $q_3$), the outcome is $x$.
If no agent in $S_3$ vote for $x$ (the number of votes for $x$ equals the quota $q_3$), the outcome is $y$.
\end{itemize}

We now identify the EMVR induced by $\Gamma _{\mathcal{S}^{o},\,Q}$. 
Observe that, throughout the process, three kinds of minimal winning coalitions for $x$ can be identified. 
The first type consists of coalitions formed as the union of a strict subset $T_1$ of $\{1,2,3\}$, with cardinality equal to $q_1=2$, and an agent $i_1 \in \{1,2,3\}\setminus T_1$.
In our example, $\{1,2,3\}$ is the unique coalition with this property.
The second type consists of coalitions formed as the union of a subset $T_1$ of $\{1,2,3\}$, with cardinality equal to $q_1=2$, a subset $T_2$ of $\{4,5,6,7,8\}$, with cardinality equal to $q_2=5$, and an agent $i_2 \in \{4,5,6,7,8\}\setminus T_2$. 
In our example, there does not exist a coalition with this property.
The third type consists of coalitions formed as the union of a subset $T_1$ of $\{1,2,3\}$, with cardinality equal to $q_1=2$, a subset $T_2$ of $\{4,5,6,7,8\}$, with cardinality equal to $q_2=5$, a subset $T_3$ of $\{9,10\}$, with cardinality equal to $q_3=0$, and an agent $i_3 \in \{9,10\}\setminus T_3$. 
In our example, these third type of coalitions are $\{1,2,4,5,6,7,8,9\}, \{1,2,4,5,6,7,8,10\}, \{1,3,4,5,6,7,8,9\}, \{1,3,4,5,6,7,8,10\}, \newline \{2,3,4,5,6,7,8,9\}$, and $\{2,3,4,5,6,7,8,10\}$. 
Therefore, the EMVR induced by $\Gamma _{\mathcal{S}^{o},\,Q}$ is the following committee for $x$:

\noindent $\mathcal{C}_m^x=\{\{1,2,3\}, \{1,2,4,5,6,7,8,9\}, \{1,2,4,5,6,7,8,10\}, \{1,3,4,5,6,7,8,9\}$

\hspace{61 pt} $\{1,3,4,5,6,7,8,10\}, \{2,3,4,5,6,7,8,9\}, \{2,3,4,5,6,7,8,10\}$.
\hfill $\square $ \medskip

In general, given any partition $\mathcal{S}$, we aim to identify all anonymous committees $\mathcal{C}^x_m$ relative to $\mathcal{S}$, whose associated EMVRs are OSP with respect to $\mathcal{S}$. 
Fix and arbitrary ordered partition $\mathcal{S}^o$ of $\mathcal{S}$ and an arbitrary compatible vector of quotas $Q$.
Define the committee of minimal winning coalitions
\vspace{-0.9cm}
\begin{center}
\begin{equation}\label{AnonCom2}
    \mathcal{C}^x_{\mathcal{S}^o,\,Q}=\bigcup _{k=1}^{K} \; \mathcal{C}^x_{\mathcal{S}^o,\,Q}(k)
\end{equation}
\end{center}
where, for each $k\in \{1,\dots, K\}$,
\vspace{-1.0cm}
\begin{center}
\begin{equation}\label{AnonCom}
\mathcal{C}^x_{\mathcal{S}^o,\,Q}(k)=\{\bigcup _{t=1}^{k}T_{t}\cup \{i_{k}\}\mid T_{t}\subset S_{t} \text{, } \left\vert T_{t}\right\vert =q_{t}\text{ and }i_{k}\in S_{k}\setminus T_{k}\}.\footnote{If $q_k=\lvert S_k\rvert$, condition (\ref{AnonCom}) implies that $\mathcal{C}^x_{\mathcal{S}^o,\,Q}(k)=\emptyset $.
This is because we cannot simultaneously predetermine a pair $T_k\subset S_k$ and $i_k\in S_k\setminus T_k$ with $\lvert T_k \rvert =q_k$.}
\end{equation}
\end{center}

To illustrate the construction of the committee $\mathcal{C}^x_{\mathcal{S}^o,\,Q}$, let's first examine the case when $k=1$ and assume $q_1<\lvert S_1\rvert$. 
In this case, we identify $\mathcal{C}^x_{\mathcal{S}^o,\,Q}(1)$ as the collection of all subsets of $N$ formed by the union $T_1\cup \{i_1\}$, where $T_1$ represents a predetermined subset of $S_1$ with cardinality equal to $q_1$, and $i_1$ denotes a predetermined agent from $S_1\setminus T_1$. 
It's important to note that there can be multiple such sets, as we can predetermine various strict subsets $T_1$ and agents $i_1 \in S_1\setminus T_1$. 
All these sets are then added to $\mathcal{C}^x_{\mathcal{S}^o,\,Q}(1)$.
If $q_1=\lvert S_1\rvert$, then, as we already argued, condition (\ref{AnonCom}) implies that $\mathcal{C}^x_{\mathcal{S}^o,\,Q}(1)=\emptyset $.

Now, let's consider the case where $1<k\leq K$  and assume $q_k<\lvert S_k\rvert$.  
In this case, we identify $\mathcal{C}^x_{\mathcal{S}^o,Q}(k)$ as the collection of all subsets of $N$ formed by the union $T_1\cup \cdots \cup T_k \cup \{i_k\}$, where for each $1\leq t \leq k$, $T_t$ represents a predetermined subset of $S_t$ with cardinality equal to $q_t$, and $i_k$ denotes a predetermined agent from the last subset $S_k\setminus T_k$. As with the previous case, there can be multiple such sets, and all of them are included in $\mathcal{C}^x_{\mathcal{S}^o,\,Q}(k)$.
If $q_k=\lvert S_k\rvert$, then, as we already argued, condition (\ref{AnonCom}) implies that $\mathcal{C}^x_{\mathcal{S}^o,\,Q}(k)=\emptyset $.
Note that the condition $q_K<\lvert S_K\rvert$, in the definition of compatibility of $Q$ with $\mathcal{S}^o$, is necessary to ensure that agents in $S_K$ are not dummies in the committee defined by (\ref{AnonCom2}).

Consequently, the committee $\mathcal{C}^x_{\mathcal{S}^o,\,Q}$ encompasses all subsets that can be obtained through this procedure. 

Now we are in a position to announce our result characterizing the class of all anonymous committees relative to a partition $\mathcal{S}$ whose associated EMVRs are OSP with respect to $\mathcal{S}$ as those satisfying condition (\ref{AnonCom2}). 

\begin{theorem}
\label{anonymity} 
Let $\mathcal{C}^{x}$ be an anonymous committee relative to a partition $\mathcal{S}$. 
Then, $f_{\mathcal{C}^{x}}:\{P^x,P^y\}^{N}\rightarrow \{x,y\}$ is OSP with respect to $\mathcal{S}$ if and only if there exist an order in $\mathcal{S}$, denoted as $\mathcal{S}^{o}=(S_{1},\dots ,S_{K})$, and a compatible vector of quotas $Q=(q_{1},\dots ,q_{K})$ such that 
\begin{equation*}
\mathcal{C}_{m}^{x}=\mathcal{C}^x_{\mathcal{S}^o,\,Q}.
\end{equation*}
 \end{theorem}

The proof of Theorem \ref{anonymity} will show that the following remark is true.

\begin{remark} Let $\mathcal{C}^{x}$ be an anonymous committee relative to a partition $\mathcal{S}$. 
Then, $f_{\mathcal{C}^{x}}:\{P^x,P^y\}^{N}\rightarrow \{x,y\}$ is OSP with respect to $\mathcal{S}$ if and only if there exist an order in $\mathcal{S}$, denoted as $\mathcal{S}^{o}=(S_{1},\dots ,S_{K})$, and a compatible vector of quotas $Q=(q_{1},\dots ,q_{K})$ such that $\Gamma _{\mathcal{S}^{o},\,Q}$ and the truth-telling type-strategy  OSP-implement $f_{\mathcal{C}^{x}}$ with respect to $\mathcal{S}$.
    
\end{remark}

The proof of Theorem \ref{anonymity} is involved.
The necessity part relies on Lemma \ref{nuevo}, which we state below. The proof of Lemma \ref{nuevo} is relegated to the Appendix in Section 5.

\begin{lemma}
\label{nuevo} 
Let $\mathcal{C}^{x}$ be an anonymous committee relative to a partition $\mathcal{S}$ and let $f_{\mathcal{C}^{x}}:\{P^x,P^y\}^{N}\rightarrow \{x,y\}$ be OSP with respect to $\mathcal{S}$.
Then, there exists an order of $\mathcal{S}$, denoted as $\mathcal{S}^{o}=(S_{1},\dots ,S_{K})$, such that, for all $k\in \{1,\dots ,K\}$, the following two conditions for $(S_1,\dots,S_k)$ hold. \newline
\emph{(i)} For all $M^\ast_k,M^{\prime }_k\in C^{x}_m$ such that $M^\ast_k\cap (\bigcup_{r=1}^{k}S_{r}) \notin C^{x}$ and $M^{\prime }_k\cap (\bigcup_{r=1}^{k}S_{r}) \notin C^{x}$, it holds that $\lvert M^\ast_k\cap S_{k}\rvert =\lvert M^{\prime }_k\cap S_{k}\rvert $.\newline
\emph{(ii)} If $M^\ast_k\in C^{x}_{m}$ is such that $M^\ast_k\cap (\bigcup_{r=1}^{k}S_{r}) \notin C^{x}$ and there exists $i\in S_{k}\setminus M^\ast_k$, then $ (M^\ast_k\cap (\bigcup_{r=1}^{k}S_{r}))\cup \{i\}\in C^{x}$.
\end{lemma}

\noindent \textbf{Proof of Theorem \ref{anonymity}. }
Let $\mathcal{C}^{x}$ be an anonymous committee relative to a partition $\mathcal{S}$. \smallskip

\noindent $(\Rightarrow )$ Suppose $\Gamma $ is such that $(\Gamma ,(\sigma _{i}^{P_{i}})_{P_i \in \mathcal{P}\,,\,i\in N})$ OSP-implements $f_{\mathcal{C}^{x}}:\{P^x,P^y\}^{N}\rightarrow \{x,y\}$ with respect to $\mathcal{S}$.

Assume $K=1$. 
By anonymity relative to $\mathcal{S}$, there exists $q\in \{1,\dots ,n\}$ such that $M\in \mathcal{C}^x_m$ if and only if $\lvert M\rvert =q$.
Then, $S_{1}=N$, $\mathcal {S}^o=(N)$, and define $Q=(q_{1})$ where $q_{1}=q-1$. 
By (\ref{AnonCom2}),
\begin{equation*}
\mathcal{C}_{m}^{x}=\{S\subset N \mid S=T\cup \{i\}, \lvert T \rvert =q-1 \text{ and } i\notin T\}=\{T\cup \{i\}\mid \lvert T\rvert =q_{1}\text{ and }i\in N\setminus T\}=\mathcal{C}_{\mathcal{S}^o,\,Q}^{x}.
\end{equation*}
Thus, the necessary condition of Theorem \ref{anonymity} holds.

Now assume $K>1$.
Let $\mathcal{S}^o=(S_1,\dots ,S_K)$ be the order of $\mathcal{S}$ given by Lemma \ref{nuevo}. 
Since there are no dummy agents, there exists $M^{\ast }\in \mathcal{C}_{m}^{x}$ such that 
\begin{equation}
   M^{\ast }\cap S_{K}\neq \emptyset .
    \label{X_K}
\end{equation}
Define, for each $t\in \{1,\dots, K\}$, $X_{t}^{\ast }:=M^{\ast }\cap S_{t}$. Then, $M^{\ast }$ can be written as $M^{\ast }=\bigcup_{t=1}^{K}X_{t}^{\ast }\in \mathcal{C}_{m}^{x}$. 
Define, for each $t\in\{1,\dots ,K-1\}$, $q_{t}=\lvert X_{t}^{\ast }\rvert $, $q_{K}=\lvert X_{K}^{\ast }\rvert -1$ and $Q=(q_{1},\dots ,q_{K})$. 
We now conclude this part of the proof of Theorem \ref{anonymity} by showing that $%
\mathcal{C}_{m}^{x}=\mathcal{C}_{\mathcal{S}^o,\,Q}^{x}$ holds.

First, we aim to show that $\mathcal{C}_{m}^{x}\subseteq \mathcal{C}_{\mathcal{S}^o,Q}^{x}$. 
Let $M\in \mathcal{C}_{m}^{x}$ be arbitrary. 
We choose $k\in \{1,\dots ,K\}$ such that $M\cap S_{k}\neq \emptyset $ and, for all $t\in \{k+1,\dots,K\}$, $M\cap S_{t}=\emptyset $. 
For every $1\leq t\leq k$, we define 
\begin{equation*}
X_{t}:=M\cap S_{t}.
\end{equation*}
As $M\cap S_{k}\neq \emptyset $, it follows that $\bigcup_{t=1}^{k-1}X_{t}\notin C^{x}$ and, by (\ref{X_K}), $\bigcup_{t=1}^{k-1}X_{t}^{\ast }\notin C^{x}$. 
Then, according to condition (i) in Lemma \ref{nuevo}, for every $t\in \{1,\dots ,k-1\}$ and $(S_1,\dots,S_t)$, we have 
\begin{equation} \label{igual}
    \lvert X_{t}\rvert=\lvert X_{t}^{\ast }\rvert=q_{t}.
\end{equation}

Now, we proceed to show that $\lvert M\cap S_{k}\rvert =q_{k}+1$ holds.
First, let's consider the case when $k=K$. 
By definition of $k$, 
we have $M\cap S_{K}\neq \emptyset $.
Then, according to condition (i) in Lemma \ref{nuevo},
we obtain $\vert M\cap S_{K}\rvert=\lvert X^{\ast }_{K}\rvert =q_{K}+1$.
Now, let's consider the case when $k<K$. 
Using (\ref{X_K}), (\ref{igual}), and the fact that $M\in\mathcal{C}^x_m$, we can deduce that 
\begin{equation}\label{q_k}
\vert M\cap S_{k}\rvert>\lvert X^{\ast }_{k}\rvert=q_k.
\end{equation}
Therefore, $\vert S_{k}\rvert>q_k$.
Furthermore, there exists $i_{k}\in S_{k}\setminus X_{k}^{\ast }$. 
By condition (ii) in Lemma \ref{nuevo}, we have
\begin{equation}
M^{\ast k}:=(\bigcup _{t=1}^{k}X_{t}^{\ast })\cup \{i_{k}\}\in \mathcal{C}%
_{m}^{x}.  \label{eq}
\end{equation}
Thus, $M^{\ast k}\cap S_{k}=X_{k}^{\ast }\cup \{i_{k}\}$.
Now, we will show that $\lvert M\cap S_{k}\rvert  \leq \lvert X_k^{\ast}\cup \{i_{k}\}\rvert = q_k +1 $, in which case, by (\ref{q_k}), we would have that $\lvert M\cap S_{k}\rvert = q_k +1 $. 
To reach a contradiction, assume that $\lvert M\cap S_{k}\rvert >\lvert X_{k}^{\ast }\cup \{i_{k}\}\rvert $.
Consider the permutation $\pi ^{\mathcal{S}}$ such that $\pi ^{\mathcal{S}}(X_{k}^{\ast }\cup \{i_{k}\})\subsetneq M\cap S_{k}$, $\pi ^{\mathcal{S}}(X_{t}^{\ast })=X_{t}$ for all $t=1,\dots ,k-1$ and $\pi ^{\mathcal{S}}(j)=j$ for all $j\notin \bigcup_{t=1}^{k} X_{t}^{\ast } \cup \{i_{k}\}$. 
Then, by (\ref{eq}), anonymity relative to $\mathcal{S}$ and the definition of $\pi ^{\mathcal{S}}$, $(\bigcup _{t=1}^{k-1}X_{t})\bigcup \pi ^{\mathcal{S}}(X_{k}^{\ast }\cup \{i_{k}\})\in \mathcal{C}_{m}^{x}$ and $(\bigcup _{t=1}^{k-1}X_{t})\bigcup \pi ^{\mathcal{S}}(X_{k}^{\ast }\cup \{i_{k}\})\subsetneq M$, contradicting the fact that $M\in \mathcal{C}_{m}^{x}$. 

Therefore, given an arbitrary $M\in \mathcal{C}^x_m$ we have shown that there exists $k\in \{1,\dots,K\}$ such that 
\[M\in\mathcal{C}^x_{\mathcal{S}^o,\,Q}(k)=\{\bigcup _{t=1}^{k}T_{t}\cup \{i_{k}\}\mid T_{t}\subset S_{t} \text{, } \left\vert T_{t}\right\vert =q_{t}\text{ and }i_{k}\in S_{k}\setminus T_{k}\} 
\]
holds.
This implies that $M\in \mathcal{C}^x_{\mathcal{S}^o,\,Q}$.

Now, we will prove that $\mathcal{C}^x_{\mathcal{S}^o,\,Q}\subseteq \mathcal{C}_{m}^{x}.$

Let $M\in \mathcal{C}^x_{\mathcal{S}^o,Q}$ be arbitrary. 
Then, there exists $k\in \{1,\dots,K\}$ such that $M=\bigcup_{t=1}^{k}T_{t}\cup \{i_{k}\}$ where $T_{t} \subsetneq   S_{t}$, $\lvert T_{t}\rvert =q_{t}$ for all $t\in \{1,\dots ,k\}$, $i_{k}\in S_{k}\setminus T_{k}$, and $q_k<\lvert S_k \rvert$. 
By definition of $Q=(q_1,\dots,q_K)$, $q_{t}=\vert T_{t}\rvert =\lvert X_{t}^{\ast }\rvert $ for all $t\in \{1,\dots ,K-1\}$ and $q_K=\lvert X^\ast _K \rvert -1$. 
Consider the permutation $\pi ^{\mathcal{S}}\in \Pi ^\mathcal{S}$ such that $\pi ^{\mathcal{S}}(X_{t}^{\ast })=T_{t}$ and $\pi ^{\mathcal{S}}(i_{k})=i_{k}$ for all $t\in\{1,\dots ,K-1\}$ if $k=K$ or all $t\in \{1,\dots ,k$\} if $k<K$.
First, assume that $k=K$. 
Then, $M\cap S_{K}\neq \emptyset $.
Therefore, according to condition (i) in Lemma \ref{nuevo}, $|M\cap S_{K}|=|M^{\ast }\cap S_{K}|$.
Then, $\pi ^{\mathcal{S}}(M)=M^{\ast }$.
Therefore, by anonymity relative to $\mathcal{S}$, $M\in \mathcal{C}_{m}^{x}$.
Second, assume $k<K$. 
According to condition (ii) in Lemma \ref{nuevo} and anonymity relative to $\mathcal{S}$, $M=\bigcup_{t=1}^k T_t \cup \{i_k\}\in \mathcal{C}_{m}^{x}$.\smallskip

\noindent $(\Leftarrow )$ Assume there exist an order in $\mathcal{S}$, denoted as $\mathcal{S}^{o}=(S_{1},\dots ,S_{K})$, and a compatible vector of quotas $Q=(q_{1},\dots ,q_{K})$ such that
\begin{equation*}
\mathcal{C}_{m}^{x}=\mathcal{C}_{\mathcal{S}^o,\,Q}^{x}.
\end{equation*}
We aim to show that $(\Gamma _{\mathcal{S}^o,\,Q}, (\sigma_i^{P_i})_{P_i \in \mathcal {P}\,,\,i\in N})$ OSP-implements $f_{\mathcal{C}^{x}}$ with respect to $\mathcal{S}$.

By Lemma \ref{OD}, $\sigma^{P_i}_i$ is obviously dominant for each $i\in N$ and each $P_i\in \{P^x,P^y\}$.

We now establish that $\Gamma _{\mathcal{S}^o,\,Q}$ and $(\sigma_i^{P_i})_{P_i \in \mathcal {P}\,,\,i\in N}$ induce $f_{\mathcal{C}^x}$ by going through the sequence of steps defining $\Gamma_{\mathcal{S}^o,\,Q}$. 
Fix an arbitrary profile $P\in \{P^x,P^y\}^N$.

\begin{itemize}

\item \underline{Step $k$} ($1\leq k<K$): 
Let $h^{k-1}$ be the history at the end of Step $k-1$ induced by $(\sigma_i^{P_i})_{P_i\in \mathcal{P}\,,\,i\in S_1\cup \cdots \cup S_{k-1}}$.
Agents in $S_{k}$ play only once and simultaneously. 
The set of available choices $Ch(I_i)$ for each $i\in S_k$ at each $I_i\in \mathcal{I}_i$ is equal to the partition $\{\{P^x\},\{P^y\}\}$.
Let $h^k$ be the history at the end of Step $k$ induced by $(\sigma_i^{P_i})_{P_i\in \mathcal{P}\,,\,i\in S_1\cup \cdots \cup S_{k-1}}$. 
We distinguish among three different cases, depending on the feature of $h^k$.

(i) $h^{k}$ is terminal and $g(z^{\Gamma _{\mathcal{S}^o,\,Q}}(z_0,\sigma^{P}))=x$. 
By the definition of $\Gamma _{\mathcal{S}^o,\,Q}$, for each $t\in \{1,\dots,k-1\}$, exactly $q_t$ agents in $S_t$ have chosen $\{P^x\}$ and strictly more than $q_k$ agents in $S_k$ have also chosen $\{P^x\}$. 
According to its definition in (\ref{AnonCom}), this set belongs to $\mathcal{C}^x_{\mathcal{S}^o,\,Q}(k)$ and, by hypothesis, to $\mathcal{C}^x$.
By (\ref{GMV}), $f_{\mathcal{C}^x}(P)=x$.
Hence, $f_{\mathcal{C}^x}(P)=g(z^{\Gamma _{\mathcal{S}^o,\,Q}}(z_0,\sigma^{P}))$.

(ii) $h^{k}$ is terminal and $g(z^{\Gamma _{\mathcal{S}^o,\,Q}}(z_0,\sigma^{P}))=y$. 
By the definition of $\Gamma _{\mathcal{S}^o,\,Q}$, for each $t\in \{1,\dots ,k-1\}$, exactly $q_t$ agents in $S_t$ have chosen $\{P^x\}$ and strictly fewer than $q_k$ agents in $S_k$ have also chosen $\{P^x\}$. 
According to its definition in (\ref{AnonCom2}), this set does not belong to $\mathcal{C}_{\mathcal{S}^o,\,Q}^{x}$ and so, by hypothesis, not all members of a winning coalition in $\mathcal{C}^x$ have chosen $\{P_i^{y}\}$. 
By (\ref{GMV}), $f_{\mathcal{C}^x}(P)=y$.
Hence, $f_{\mathcal{C}^x}(P)=g(z^{\Gamma _{\mathcal{S}^o,\,Q}}(z_0,\sigma^{P}))$.

(iii) $h^{k}$ is non-terminal. By the definition of $\Gamma_{\mathcal{S}^o,\,Q}$, for each $1\leq t\leq k$, exactly $q_t$ agents in $S_t$ have chosen $\{P^x\}$.
According to the definition of $\Gamma_{\mathcal{S}^o,\,Q}$, the $[\mathcal{S}^o,Q]$-process proceeds to Stage $k+1$ with $h^k$.

\item \underline{Step $K$}: 
Let $h^{K-1}$ be the history at the end of Step $K-1$ induced by $(\sigma_i^{P_i})_{P_i \in \mathcal{P}\,,\,i\in N\setminus S_{K}}$.
Agents in $S_{K}$ play only once and simultaneously. 
The set of available choices $Ch(I_i)$ for each $I_i\in\mathcal{I}_i$ is equal to the partition $\{\{P^x\},\{P^y\}\}$. 
Let $h^K$ be the history at the end of Step $K$ induced by $(\sigma_i^{P_i})_{P_i \in \mathcal{P}\,,\,i\in N}$. 
Since $K$ is the last step of the $[\mathcal{S}^o,Q]$-process, $h^K$ is terminal.
We distinguish between two different cases, depending on the outcome associated to $h^K$.

(i) $g(z^{\Gamma _{\mathcal{S}^o,\,Q}}(z_0,\sigma^{P}))=x$.
By the definition of $\Gamma _{\mathcal{S}^o,\,Q}$, for each $t\in \{1,\dots K-1\}$, exactly $q_t$ agents in $S_t$ have chosen $\{P^x\}$ and strictly more than $q_K$ agents in $S_K$ have also chosen $\{P^x\}$.
According to its definition in (\ref{AnonCom}), this set belongs to $\mathcal{C}^x_Q(K)$ and, by hypothesis, to $\mathcal{C}^x$.
By (\ref{GMV}), $f_{\mathcal{C}^x}(P)=x$.
Hence, $f_{\mathcal{C}^x}(P)=g(z^{\Gamma _{\mathcal{S}^o,\,Q}}(z_0,\sigma^{P}))$.

(ii) $g(z^{\Gamma _{\mathcal{S}^o,\,Q}}(z_0,\sigma^{P}))=y$. 
By the definition of $\Gamma _{\mathcal{S}^o,\,Q}$, for each $t\in \{1\dots K-1\}$, exactly $q_t$ agents in $S_t$ have chosen $\{P^x\}$ and fewer than or equal to $q_K$ agents in $S_K$ have also chosen $\{P^x\}$.
According to its definition in (\ref{AnonCom2}), this set does not belong to $\mathcal{C}_{\mathcal{S}^o,\,Q}^{x}$ and so, by hypothesis, not all members of a winning coalition in $\mathcal{C}^x$ have chosen $\{P^{x}\}$. 
By (\ref{GMV}), $f_{\mathcal{C}^x}(P)=y$.
Hence,  $f_{\mathcal{C}^x}(P)=g(z^{\Gamma _{\mathcal{S}^o,\,Q}}(z_0,\sigma^{P}))$.
\end{itemize}

Therefore, $f_{\mathcal{C}^x}(P)=g(z^{\Gamma _{\mathcal{S}^o,\,Q}}(z_{0},\sigma ^{P}))$.

Thus, the game $\Gamma _{\mathcal{S}^o,\,Q}\in \mathcal{F}^{\mathcal{S}^{o},\,Q}$ and the truth-telling type strategy profile $(\sigma_i^{P_i})_{P_i \in \mathcal{P}\,,\,i\in N}$ OSP-implement $f_{\mathcal{C}^{x}}$ with respect to $\mathcal{S}$. 
This finishes the proof of Theorem \ref{anonymity}.\hfill $\blacksquare $

\subsubsection{Strong anonymity}
A committee $\mathcal{C}^{x}$ is \emph{strongly anonymous} if for all bijections $\pi:N\rightarrow N$, $M\in \mathcal{C}^{x}$ if and only if $\pi (M)\in \mathcal{C}^{x}$.
This is the straightforward definition of an anonymous committee that does not take into account the partition. 
Of course, any strongly anonymous committee is anonymous with respect to any partition.

\begin{remark}
\label{q} Let $\mathcal{C}^{x}$ be strongly anonymous.
Then, there exists an integer $q\in \{1,\dots ,n\}$, called the \emph{quota} associated with $\mathcal{C}^x$, such that, $M\in \mathcal{C}_{m}^{x}$ if and only if $\lvert M\rvert =q$.\hfill $\square$ 
\end{remark}

\begin{theorem}\label{SA} 
Let $\mathcal{C}^{x}$ be a strongly anonymous committee with associated quota $q$ and let $\mathcal{S}$ be a partition. 
Then, $f_{\mathcal{C}^{x}}:\{P^x,P^y\}^{N}\rightarrow \{x,y\}$ is OSP with respect to $\mathcal{S}$ if and only if at least one of the following three statements hold:
\newline
\textnormal{(i)} $q\in \{1,n\}$. 
\newline
\textnormal{(ii)} $K=1$. \newline
\textnormal{(iii)} $K=2$ and $\lvert S_{1}\rvert \in \{1,n-1\}$.\footnote{Note that conditions (i), (ii) and (iii) are not mutually exclusive.\smallskip}
\end{theorem}

\noindent \textbf{Proof of Theorem \ref{SA}. } Let $\mathcal{C}^{x}$ be a strongly anonymous committee with associated quota $q$ and let $\mathcal{S}$ be a partition. 
Then, $\mathcal{C}^c$ is anonymous with respect to $\mathcal{S}$.\smallskip

\noindent $(\Rightarrow )$ Assume $f_{\mathcal{C}^{x}}:\{P^x,P^y\}^{N}\rightarrow \{x,y\}$ is OSP with respect to $\mathcal{S}$. 
By Theorem \ref{anonymity}, there exist an order in $\mathcal{S}$, denoted as $\mathcal{S}^{o}=(S_{1},\dots ,S_{K})$, and a compatible vector of quotas $Q=(q_{1},\dots ,q_{K})$ such that 
\begin{equation}
\mathcal{C}_{m}^{x}=\mathcal{C}_{\mathcal{S}^o,\,Q}^{x}.  \label{4}
\end{equation}

We proceed by distinguishing among three exhaustive but now mutually exclusive cases, which are determined by the value of $K$, representing the number of subsets in $\mathcal{S}$.\smallskip

\noindent Case 1: $K=n$; \textit{i.e.}, $\mathcal{S}=\{\{1\},\dots ,\{n\}\}$.
By (R1.2) in Remark \ref{RemK=1,K=n}, $f_{\mathcal{C}^{x}}$ is OSP. 
By Corollary 2 in Arribillaga, Massó and Neme (2020), $q\in \{1,n\}$. 
Thus, the conclusion of condition (i) holds for this case.\smallskip

\noindent Case 2: $K=1$.
Condition (ii) holds trivially.\smallskip

\noindent Case 3: $1<K<n$. 
Suppose that neither (i) nor (ii) in the statement of the theorem holds. 
We now prove that condition (iii) holds.
Suppose there exists $k \in\{1,\dots,K\}$ such that $1<\lvert S_{k}\rvert <n-1$. 
Then, by strong anonymity, there exists $M\in \mathcal{C}_{m}^{x}$ such that $\lvert M\cap S_{k}\rvert <q$. 
Since $\mathcal{C}_{m}^{x}$ is strongly anonymous, for all $M^{\prime }\subseteq N$ such that $\lvert M^{\prime }\rvert =q$, $M^{\prime }\in \mathcal{C}_{m}^{x}$.
Moreover, as $1<\lvert N\setminus S_{k}\rvert $, we can find $j^{\prime }\notin S_{k}\cup M$ and $j\in M\cap S_{k}$. 
Define $M^{\prime }=(M\setminus \{j\})\cup \{j^{\prime }\}$. 
Then, $\lvert M^{\prime }\rvert =q$, and so $M^{\prime }\in \mathcal{C}_{m}^{x}$.
Therefore, there are two minimal winning coalitions $M,M^{\prime }\in \mathcal{C}_{m}^{x}$ such that $M\cap S_{k},M^{\prime }\cap S_{k}\notin \mathcal{C}^{x}$ and $\lvert M\cap S_{k}\rvert >\lvert M^{\prime }\cap S_{k}\rvert $, which contradicts condition (i) in Lemma \ref{nuevo}.
Then, for all $k\in \{1,\dots, K\}$, either $\lvert S_{k}\rvert =1$ or $\lvert S_{k}\rvert =n-1$, which implies that $K=2$ and $\lvert S_{1}\rvert \in \left\{ 1,n-1\right\}$.
\medskip

\noindent ($\Leftarrow )$ Assume (i) holds.
Then, the sufficient condition of the theorem follows from (R1.2) in Remark \ref{RemK=1,K=n} and Corollary 2 in Arribillaga, Massó and Neme (2020).

Assume (ii) holds.
Then, the sufficient condition of the theorem follows from (R1.1) in Remark \ref{RemK=1,K=n} and Corollary 1 in Barberà, Sonnenschein and Zhou (1991).

Assume (iii) holds and, without loss of generality, suppose $\lvert S_1\rvert=n-1$.
Consider the ordered partition $\mathcal{S}^o=(S_1,S_2)$ and the compatible vector of quotas $Q=(q_1,q_2)$, where $q_1=q-1$ and $q_2=0$.
Observe that $q_1\leq\lvert S_1\rvert$ and $q_2<\lvert S_2\rvert$.
If $q_1=\lvert S_1\rvert$, then $q=n$ and the proof follows as in case (i).
If $q_1<\lvert S_1\rvert$, then $\mathcal{C}_{\mathcal{S}^o,\,Q}^{x}(1)=\{T_1\cup\{i_1\}\mid T_1\subset S_1, \lvert S_1\rvert =q_1 \text { and } i_1\in S_1\setminus T_1\}=\mathcal{C}_{\mathcal{S}^o,\,Q}^{x}(2)$ .
Hence, $\mathcal{C}^x_m=\mathcal{C}_{\mathcal{S}^o,\,Q}^{x}$.
Therefore, the necessary condition in Theorem \ref{anonymity} holds and, accordingly, $f_{\mathcal{C}^{x}}:\{P^x,P^y\}^{N}\rightarrow \{x,y\}$ is OSP with respect to $\mathcal{S}$.
Then, the sufficient condition of the theorem holds.\hfill $\blacksquare $

\section{Appendix: Proof of Lemma \ref{nuevo}}

\setcounter{lemma}{1}

\begin{lemma}  
Let $\mathcal{C}^{x}$ be an anonymous committee relative to a partition $\mathcal{S}$ and let $f_{\mathcal{C}^{x}}:\{P^x,P^y\}^{N}\rightarrow \{x,y\}$ be OSP with respect to $\mathcal{S}$.
Then, there exists an order of $\mathcal{S}$, denoted as $\mathcal{S}^{o}=(S_{1},\dots ,S_{K})$, such that, for all $k\in \{1,\dots ,K\}$, the following two conditions for $(S_1,\dots,S_k)$ hold. \newline
\emph{(i)} For all $M^\ast_k,M^{\prime }_k\in C^{x}_m$ such that $M^\ast_k\cap (\bigcup_{r=1}^{k}S_{r}) \notin C^{x}$ and $M^{\prime }_k\cap (\bigcup_{r=1}^{k}S_{r}) \notin C^{x}$, it holds that $\lvert M^\ast_k\cap S_{k}\rvert =\lvert M^{\prime }_k\cap S_{k}\rvert $.\newline
\emph{(ii)} If $M^\ast_k\in C^{x}_{m}$ is such that $M^\ast_k\cap (\bigcup_{r=1}^{k}S_{r}) \notin C^{x}$ and there exists $i\in S_{k}\setminus M^\ast_k$, then $ (M^\ast_k\cap (\bigcup_{r=1}^{k}S_{r}))\cup \{i\}\in C^{x}$.
\end{lemma}

\noindent \textbf{Proof of Lemma \ref{nuevo}. } Let $\mathcal{C}^{x}$ be an anonymous committee relative to a partition $\mathcal{S}$, and let $\Gamma $ be the extensive game form that OSP-implements $f_{\mathcal{C}^{x}}$ with respect to $\mathcal{S}$. 
By anonymity, $f_{\mathcal{C}^{x}}$ does not have dummy agents. 
Thus, according to Remark \ref{RTM} in Subsection 3.3, we can assume that $\Gamma $ is a round table mechanism. 
Moreover, any complete history $h\in H$ can be written as $h=z_{0},\dots ,z_{\ell ^{\prime }},\dots ,z_{\ell }$, where $\ell \leq n$, $z_{\ell }\in Z_{T}$ is a terminal node, each agent $i$ plays at most once along $h$, and for every $\ell^\prime\in \{0,\dots ,\ell -1\}$, $Ch(z_{\ell^\prime})=\{\{P^{x}\},\{P^{y}\}\}$. 
For any $\ell ^{\prime }\in \{0,\dots ,\ell \}$, we denote the agent who owns node $z_{\ell ^{\prime }}$ as $\mathcal{N}(z_{\ell^{\prime }})$; that is, $z_{\ell ^{\prime }}\in Z_{\mathcal{N}(z_{\ell^{\prime }})}$.

Assume $K=1$. 
In this case, $S_1=N$ and the statement of Lemma \ref{nuevo} follows trivially because the conditions in (i) and (ii) do not apply.

Assume $K>1$. 
The proof proceeds by induction on $k$.\footnote{
Note that for $1\leq k < K$, the conditions (i) and (ii) in Lemma \ref{nuevo} only require the identification of the partial ordered partition $(S_{1},\dots ,S_{k})$, rather than the complete ordered partition $(S_{1},\dots ,S_{K})$.\smallskip}

First, set $k=1$. 
Let $i_{1}\in N$ be the first agent to play in $\Gamma $ (\textit{i.e.}, $\mathcal{N}(z_{0})=i_{1}$) by choosing, at $i_{1}$'s unique information set $I_{i_{1}}=\{z_{0}\}$, one from the set of choices $Ch(I_{i_{1}})=\{\{P^{x}\},\{P^{y}\}\}$. 
Let $S_{1}$ be the element of $\mathcal{S}$ that contains $i_{1}$. 
The ordered partition $\mathcal{S}^{o}$, whose existence we have to show, begins with $S_{1}$. 
Since there are no dummy agents and $k=1<K$, there exists at least one minimal winning coalition $M_{1}\in \mathcal{C}_{m}^{x}$ such that $M_{1}\cap S_{1}\notin \mathcal{C}_{m}^{x}$.

To prove that condition (i) holds for $k=1$ and $S_1$ by contradiction, assume there exist $M_{1}^{\ast },M_{1}^{\prime }\in \mathcal{C}_{m}^{x}$ such that $X_{1}^{\ast}:=M_{1}^{\ast }\cap S_{1}\notin \mathcal{C}^{x}$, $X_{1}^{\prime}:=M_{1}^{\prime }\cap S_{1}\notin \mathcal{C}_{m}^{x}$, and $\left\vert X_{1}^{\ast }\right\vert \neq \left\vert X_{1}^{\prime }\right\vert $.
Assume $\left\vert X_{1}^{\ast }\right\vert <\left\vert X_{1}^{\prime}\right\vert $. 
By anonymity of $\mathcal{C}^x$ relative to $\mathcal{S}$, we can assume without loss of generality that $X_{1}^{\ast }\subsetneq X_{1}^{\prime }$ and $i_{1}\in X_{1}^{\prime }\setminus X_{1}^{\ast }$. 
For each $j\in N\setminus \{i_{1}\}$, let $\mathcal{I}_{j}$ be the family of information sets where agent $j$ has to make a choice in $\Gamma $, with the set of choices at those information sets being $\{\{P^{x}\},\{P^{y}\}\}$. 
Since $\mathcal{C}^{x}$ has no dummy agents, each agent $j\in N\setminus \{i_{1}\}$ has at lest one information set with this property (\textit{i.e.}, $\mathcal{I}_{j}\neq \emptyset $). 

Assume that agent $i_{1}$'s preference is $P_{i_{1}}=P^{x}$. 
Consider the profile $P_{-i_1}=(P_{S_{1}\setminus \{i_{1}\}},P_{N\setminus S_{1}})$ where $P_{j}=P^{x}$ for all $j\in X_{1}^{\ast }$, $P_{j}=P^{y}$ for all $j\in S_{1}\setminus (X_{1}^{\ast }\cup \{i_{1}\})$, and $P_{j}=P^{y}$ for all if $j\in N\setminus S_{1}$. 
Define $P=(P _{i_1},P _{S_1\setminus \{i_1\}},P _{N\setminus S_{1}})$ and observe that $\{i\in N \mid P_i=P^x\}=X_{1}^{\ast }\cup \{i_{1}\}\subseteq X^\prime _1\notin \mathcal{C}^{x}$.
Therefore, by the definition of $f_{\mathcal{C}^x}$ in (\ref{GMV}), we have  
\begin{equation}\label{f(R)=y}
f_{\mathcal{C}^x}(P)=y.
\end{equation}
Let $\sigma^P=(\sigma _{i_{1}}^{P_{i_{1}}},\sigma^P _{S_{1}\setminus \{i_{1}\}},\sigma^P _{N\setminus S_{1}})$ be the truth-telling strategy associated with $P$. 
Specifically, $\sigma_{i_{1}}^{P_{i_{1}}}(z_0)=\{P^x\}$, $\sigma_{j}^{P_j}(I_j)=\{P^x\}$ for all $j\in X_1^\ast$ and all $I_j\in \mathcal{I}_j$, $\sigma_j^{P_j}(I_j)=\{P^y\}$ for all $j\in S_1\setminus (X_1^\ast \cup \{i_1\})$ and all $I_j\in \mathcal{I}_j$, and $\sigma_j^{P_j}(I_j)=\{P^y\}$ for all $j\in N\setminus S_1$ and all $I_j\in \mathcal{I}_j$.
Since $\Gamma $ and the type-strategy profile induce $f_{\mathcal{C}^{x}}$, it follows that
\begin{equation}\label{f=g}
f_{\mathcal{C}^{x}}(P)=g(z^{\Gamma }(z_{0},(\sigma _{i_{1}}^{P_{i_{1}}},\sigma^P _{S_{1}\setminus \{i_{1}\}},\sigma^P _{N\setminus S_{1}}))).
\end{equation}
Combining conditions (\ref{f(R)=y}) and (\ref{f=g}), we obtain
\begin{equation*}
y=g(z^{\Gamma }(z_{0},(\sigma _{i_{1}}^{P_{i_{1}}},\sigma^P _{S_{1}\setminus \{i_{1}\}},\sigma^P _{N\setminus S_{1}}))).
\end{equation*}
Thus, $y$ is the outcome when $i_1$ chooses $\{P^x\}$ while the remaining agents follow their strategies $\sigma_{-i}^P$.

Now, assume that agent $i_{1}$ deviates at $I_{i_{1}}=\{z_{0}\}$ with the truth-telling strategy corresponding to the preference $P^\prime_{i_1}=P^y$; \textit{i.e.}, $\sigma^{P'_{i_{1}}}_{i_{1}}(z_0)=\{P^y\}$.
Therefore, the relevant (and trivial) earliest points of departure for $\sigma^{P} _{S_{1}}$ and $\sigma^{P_{i_1}^\prime}_{i_{1}}$ is $I_{i_{1}}(\sigma _{S_{1}},\sigma^{P^\prime_{i_1}}_{i_{1}})=\{z_{0}\}$. 
Let $P^\prime_{N\setminus S_{1}}$ be the preference profile for $N\setminus S_{1}$ such that $P'_{j}=P^{x}$ for all $j\in N\setminus S_{1}$.
Define $P^\prime=(P^\prime _{i_1},P _{S_1\setminus \{i_1\}},P^\prime _{N\setminus S_{1}})$ and let $\sigma^{P^\prime} _{N\setminus S_{1}}$ be the truth-telling strategy associated with $P^\prime_{N\setminus S_{1}}$.
Observe that $\{i\in N \mid P^\prime_i=P^x\}=X^{\ast }_1\cup (N\setminus S_{1})$.    
Then, as $M^{\ast }_1\subset X^{\ast }_1\cup (N\setminus S_{1})$ and $M^{\ast }_1\in \mathcal{C}^x$, the monotonicity of $\mathcal{C}^x$ implies that $X^{\ast }_1\cup (N\setminus S_{1})\in \mathcal{C}^x$.
Therefore, by the definition of $f_{\mathcal{C}^x}$ in (\ref{GMV}), we have
\begin{equation}\label{f(P')=x}
    f_{\mathcal{C}^{x}}(P^\prime)=x.
\end{equation}
Since $\Gamma $ and the type-strategy profile induce $f_{\mathcal{C}^{x}}$, it follows that
\begin{equation}\label{f(P')=g}
    f_{\mathcal{C}^{x}}(P^\prime)=g(z^{\Gamma }(z_{0},(\sigma^{P'_{i_1}}_{i_{1}},\sigma^P _{S_{1}\setminus\{i_{1}\}},\sigma^{P'}_{N\setminus S_{1}}))).
\end{equation}
Combining conditions (\ref{f(P')=x}) and (\ref{f(P')=g}), we obtain
\begin{equation}
x=g(z^{\Gamma }(z_{0},(\sigma^{P'_{i_1}}_{i_{1}},\sigma^P _{S_{1}\setminus\{i_{1}\}},\sigma^{P'}_{N\setminus S_{1}}))).
\end{equation}
Thus, $x$ is the outcome when $i_1$ chooses $\{P^y\}$ while the remaining agents follow their strategies $\sigma_{-i}^{P^\prime}$.
Since $y\in o(\sigma^P _{S_{1}},\sigma^{P'_{i_1}}_{i_{1}})$, $x\in o^{\prime }(\sigma^P _{S_{,1}},\sigma^{P'_{i_1}}_{i_{1}})$ and $x\,P_{i_{1}}\,y$, we have that $\sigma _{i_{1}}^{P_{i_{1}}}$ is not obviously dominant with respect to $\mathcal{S}$ in $\Gamma $ for $i_{1}$ with $P_{i_{1}}=P^{x}$; a contradiction with the hypothesis that $(\Gamma,(\sigma_i^{P_i})_{P_i \in \mathcal{P}\,,\,i\in N})$ OSP-implements $f_{\mathcal{C}^{x}}$ with respect to $\mathcal{S}$.
Proceed similarly to obtain a contradiction for the other case where $\left\vert X_{1}^{\ast }\right\vert >\left\vert X_{1}^{\prime }\right\vert $.
Hence, the necessary condition (i) of Lemma \ref{nuevo} holds for $k=1$ and $S_1$.

To prove that condition (ii) holds for $k=1$ and $S_1$ by contradiction, assume there exists $M_{1}^{\ast}\in C_{m}^{x}$ such that $X_{1}^{\ast }:=M_{1}^{\ast }\cap S_{1}\notin \mathcal{C}^{x}$, and for all $j\in S_{1}\setminus M_{1}^{\ast }$, 
\begin{equation*}
X_{1}^{\ast }\cup \{j\}\notin \mathcal{C}^{x}.
\end{equation*}
Let $j$ be an agent with this  property.
By anonymity relative to $\mathcal{S}$, we can swap agent $j$ with agent $i_1$ identified in the proof of part (i).
The proof of condition (ii) then follows the same arguments used in the proof of (i), starting from the assumption that agent $i_1$'s preference is $P_{i_1}=P^x$.
Thus, the necessary condition (ii) of Lemma \ref{nuevo} holds for $k=1$ and $S_1$.

Let $1\leq k<K$ and assume that the necessary conditions (i) and (ii) of Lemma \ref{nuevo} hold for each $t=1,\dots ,k$ for $(S_1,\dots,S_t)$, where $(S_1,\dots, S_k)$ is the order of elements of $\mathcal{S}$ identified along the inductive proof up to $k$. 

We shall identify $S_{k+1}\in \mathcal{S}\setminus \{S_1,\dots,S_k\}$ and show that the necessary conditions (i) and (ii) of Lemma \ref{nuevo} hold as well for $k+1$ and $(S_1,\dots,S_k,S_{k+1})$.

Since there are no dummy agents and $k<K$, there exists $\widehat{M}\in C_{m}^{x}$ such that 
\begin{equation}
\widehat{M}\cap (\textstyle \bigcup\limits_{r=1}^{k}S_{r})\notin C_{m}^{x}.  \label{a}
\end{equation}

Fix the preference profile $P\in \mathcal{P}^N$, where $P_j=\{P^x\}$ for all $j\in \widehat{M}$ and $P_j=\{P^y\}$ for all $j\notin \widehat{M}$.
Therefore, by the definition of $f_{\mathcal{C}^x}$ in (\ref{GMV}), we have
\begin{equation}\label{f(P)=xbis}
f_{\mathcal{C}^x}(P)=x.
\end{equation}

Let $\sigma^P$ be the truth-telling strategy profile where $\sigma^{P_j} _{j}(I_{j})=\{P^{x}\}$ for all $j\in \widehat{M}$ and all $I_{j}\in \mathcal{I}_{j}$, and $\sigma^{P_j} _{j}(I_{j})=\{P^{y}\}$ for all $j\in N\setminus \widehat{M}$ and all $I_{j}\in \mathcal{I}_{j}$. 
Since $\Gamma$ and the type-strategy profile induce $f_{\mathcal{C}^x}$,
\begin{equation}\label{f=gbis}
   f_{\mathcal{C}^x}(P)= g(z^\Gamma(z_0,\sigma^P)).
\end{equation}
Combining conditions (\ref{f(P)=xbis}) and (\ref{f=gbis}), we obtain
\begin{equation*}
    x=g(z^\Gamma(z_0,\sigma^P)).
\end{equation*}
Let $h=z_{0},\ldots z_{\ell ^{\prime }},\ldots ,z^{\Gamma }(z_0,\sigma^P)$ be the complete history induced by $\sigma^P$.
Let $N(h)$ be the set of agents that play along $h$, and among them, let $N(h^x)$ and $N(h^y)$ be those that play $\{P^x\}$ and $\{P^y\}$, respectively.
Notice that 
\begin{equation} \label{JAA}
N(h^{x})\subseteq \widehat{M}=\{i\in N\mid P_i=\{P^x\}\}
\end{equation}
and 
\begin{equation}\label{JAAbis}
N(h^{y})\subseteq N\setminus \widehat{M}.
\end{equation}

By (\ref{a}), there exists $j\in \widehat{M}$ such that $j\notin \bigcup_{r=1}^{k}S_{r}$. 
We now prove that $j\in N(h)$. 
To obtain a contradiction, suppose $j\notin N(h)$. 
Consider the preference profile $\hat{P}$ where $\hat{P}_j=\{P^y\}$ and $\hat{P}_i=P_i$ for all $i\neq j$.
Since $\widehat{M}$ is a minimal winning coalition, $\{i\in N\mid\hat{P}_i=\{P^x\}\}=\widehat{M}\setminus\{j\}\notin \mathcal{C}^x$. 
Therefore, by the definition of $f_{\mathcal{C}^x}$ in  (\ref{GMV}), we have
\begin{equation}\label{f(P^)=y}
    f_{\mathcal{C}^x}(\hat{P})=y.
\end{equation}
Let $\sigma^{\hat{P}}=(\sigma_j^{\hat{P}_{j}},\sigma_{\widehat{M}\setminus\{j\}}^{\hat{P}},\sigma_{N\setminus \widehat{M}}^{\hat{P}})$ be the truth-telling strategy associated with $\hat{P}$. 
Since $\Gamma$ and the type-strategy profile induce $f_{\mathcal{C}^x}$,
\begin{equation}\label{f(P^)=g}
f_{\mathcal{C}^x}(\hat{P})=g(z^\Gamma(z_0,\sigma^{\hat{P}})).
\end{equation}
Combining conditions (\ref{f(P^)=y}) and (\ref{f(P^)=g}), we obtain $y=g(z^\Gamma(z_0,\sigma^{\hat{P}}))$.
Since $j\notin N(h)$ and $\sigma_i^{P_i}=\sigma_i^{\hat{P}_i}$ for all $i\neq j$,
$z^\Gamma(z_0,\sigma^P)=z^\Gamma(z_0,\sigma^{\hat{P}})$.
Hence, $x=g(z^\Gamma(z_0,\sigma^P))=g(z^\Gamma(z_0,\sigma^{\hat{P}}))=y$ which is a contradiction. 
Therefore, $j\in N(h)$.  

Let $\ell $ indicate the index for which $z_{\ell}< z^\Gamma (z_0,\sigma^P)$ has the property that all agents playing just before $z_{\ell }$ along $h$ belong to $\bigcup_{r=1}^{k}S_{r}$ and the agent playing at $z_{\ell }$ is the first one that does not belong to $\bigcup_{r=1}^{k}S_{r}$.
Let $S_{k+1}$ be such that $\mathcal{N}(z_{\ell })\in S_{k+1}$.
Namely, for all $\ell ^{\prime}\in \{0,\dots ,\ell -1\}$, $\mathcal{N}(z_{\ell ^{\prime }})\in \bigcup_{r=1}^{k}S_{r}$ and $i:=\mathcal{N}(z_{\ell })\in S_{k+1}$.

For each $\ell ^{\prime }\in \{0,\dots ,\ell -1\}$, we use the identification between $z_{\ell ^{\prime }+1}$ in $h_{\ell}=z_{0},\dots ,z_{\ell ^{\prime }},\dots ,z_{\ell }$ and the choice made by agent $\mathcal{N}(z_{\ell ^{\prime }})$ at $z_{\ell ^{\prime }}$ from the set $Ch(z_{\ell ^{\prime }})=\{\{P^{x}\},\{P^{y}\}\}$. 
Denote this choice by $a_{\mathcal{N}(z_{\ell ^{\prime }})}$ (\textit{i.e.}, the identification is between $z_{\ell ^{\prime }+1}$ and $a_{\mathcal{N}(z_{\ell ^{\prime }})}$).
Denote by 
\begin{equation*}
N(h_{\ell })=\{j\in N\mid j=\mathcal{N}(z_{\ell ^{\prime }})\text{ for some }\ell ^{\prime }\in \{0,\dots ,\ell -1\}\}
\end{equation*}
the set of agents that play along $h_{\ell}\setminus \{z_{\ell }\}$ and by 
\begin{equation*}
N(h_{\ell }^{x})=\{j\in N(h_{\ell })\mid a_{j}=\{P^{x}\}\}\text{ and }N(h_{\ell }^{y})=\{j\in N(h_{\ell })\mid a_{j}=\{P^{y}\}\}
\end{equation*}
the sets of agents that play along $h_{\ell }$ and choose $\{P^{x}\}$ and $\{P^{y}\}$, respectively. 
Observe that
\begin{equation}\label{b}
N(h_{\ell }^{x})\subseteq \widehat{M}\cap (\textstyle \bigcup_{r=1}^{k}S_{r}).
\end{equation}

To prove that condition (i) holds for $k+1$ and $(S_{1},\dots ,S_{k+1})$ by contradiction, assume there exist $M_{k+1}^{\ast },M_{k+1}^{\prime }\in \mathcal{C}_{m}^{x}$ such that $M_{k+1}^{\ast }\cap (\bigcup_{r=1}^{k+1}S_{r})\notin \mathcal{C}^{x}$, $M_{k+1}^{\prime }\cap (\bigcup_{r=1}^{k+1}S_{r})\notin C^{x}$, $X_{k+1}^{\ast }:=M_{k+1}^{\ast }\cap S_{k+1}$, $X_{k+1}^{\prime}:=M_{k+1}^{\prime }\cap S_{k+1}$, and $\lvert X_{k+1}^{\ast }\rvert \neq \lvert X_{k+1}^{\prime }\rvert $. 
Suppose that $\lvert X_{k+1}^{\ast }\rvert<\left\vert X_{k+1}^{\prime }\right\vert $. 
By anonymity of $\mathcal{C}^x$ relative to $\mathcal{S}$, we can assume without loss of generality that $X_{k+1}^{\ast }\subsetneq X_{k+1}^{\prime} $ and $i\in X_{k+1}^{\prime }\setminus X_{k+1}^{\ast }$.

By the induction hypothesis and anonymity of $\mathcal{C}^x$ relative to $\mathcal{S}$ we can assume that 
\begin{equation}\label{igualdad}
\widehat{M}\cap (\textstyle \bigcup_{r=1}^{k}S_{r})=M_{k+1}^{\ast }\cap (\textstyle \bigcup_{r=1}^{k}S_{r})=M_{k+1}^{\prime }\cap (\textstyle \bigcup_{r=1}^{k}S_{r}),  
\end{equation}
where $\widehat{M}\in\mathcal{C}^ x$ was identified in (\ref{a}).

For each $j\in N(h)$, let $\mathcal{I}_{j}$ be the family of agent $j$'s information sets where $j$ has to choose in $\Gamma $ and the set of choices at those information sets is $\{\{P^{x}\},\{P^{y}\}\}$. Since $\mathcal{C}^{x}$ has no dummy agents, each agent $j$ that plays along $h$ up to node $z_{\ell }$ (\textit{i.e.}, $j\in N(h_\ell)$) has at least one information set with this property.

Assume that agent $i$'s preference is $P_{i}=\{P^{x}\}$. 
Consider the profile $P_{-i}=(P_{S_{k+1}\setminus \{i\}},P_{N\setminus S_{k+1}})$ where $P_j=\{P^x\}$ for all $j\in X^\ast_{k+1}$, $P_j=\{P^y\}$ for all $j\in S_{k+1}\setminus (X^\ast_{k+1}\cup\{i\})$, $P_j=\{P^x\}$ for all $j\in N(h^x_\ell)$, and $P_j=\{P^y\}$ for all $j\in N\setminus (S_{k+1}\cup N(h^x_\ell))$.
Define $P=(P_i,P_{S_{k+1}\setminus \{i\}},P_{N\setminus S_{k+1}})$ and observe that $\{j\in N\mid P_j=\{P^x\}\}=X^\ast_{k+1}\cup \{i\}\cup N(h^x_\ell)\}$.
By (\ref{b}) and the induction hypothesis, $X^\ast_{k+1}\cup \{i\}\cup N(h^x_\ell)\}\notin \mathcal{C}^x$. 
Therefore, by the definition of $\mathcal{C}^x$ in (\ref{GMV}), we have
\begin{equation}\label{f(P)=y 2}
f_{\mathcal{C}^x}(P)=y.
\end{equation}
Let $\sigma ^P=(\sigma_i^{P_i},\sigma_{S_{k+1}\setminus\{i\}}^P,\sigma_{N\setminus S_{k+1}}^P)$ be the truth-telling strategy associated with $P$.
Specifically, $\sigma ^{P_i}_i(I_i)=\{P^x$\} for all $I_i\in \mathcal{I}_i$, $\sigma ^{P_j}_i(I_j)=\{P^x\}$ for all $j\in X^\ast _{k+1}$ and all $I_j\in \mathcal{I}_j$, $\sigma ^{P_j}_j(I_j)=\{P^y\}$ for all $j\in S_{k+1}\setminus (X^\ast_{k+1}\cup \{i\})$ and all $I_j\in \mathcal{I}_j$, $\sigma^{P_j}_j(I_j)=\{P^x\}$ for all $j\in N(h^x_\ell)$ and all $I_j\in \mathcal{I}_j$, and $\sigma^{P_j}_j(I_j)=\{P^y\}$ for all $j\in N\setminus (S_{k+1}\cup N(h^x_\ell))$ and all $I_j\in \mathcal{I}_j$.
Since $\Gamma $ and the type-strategy profile induce $f_{\mathcal{C}^x}$, it follows that
\begin{equation}\label{f(P)=y bis}
f_{\mathcal{C}^x}(P)=g(z^{\Gamma }(z_{0},(\sigma _{i}^{P_{i}},\sigma _{S_{k+1}\setminus \{i\}}^{P},\sigma _{N\setminus S_{k+1}}^{P}))).
\end{equation}      
Combining conditions (\ref{f(P)=y 2}) and (\ref{f(P)=y bis}), we obtain
\begin{equation*}
y=g(z^{\Gamma }(z_{0},(\sigma _{i}^{P_{i}},\sigma _{S_{k+1}\setminus \{i\}}^{P},\sigma _{N\setminus S_{k+1}}^{P}))).
\end{equation*}      
Since, by definition of $P$, $z_{\ell }\in Z_{i}$ belongs to the path induced
by the strategy profile $P$, we can express this as 
\begin{equation*}
y=g(z^{\Gamma }(z_{\ell},(\sigma _{i}^{P_{i}},\sigma _{S_{k+1}\setminus
\{i\}}^{P},\sigma _{N\setminus S_{k+1}}^{P}))).
\end{equation*}
Thus, $y$ is the outcome, starting at $z_\ell$, when $i$ chooses $\{P^x\}$ while the remaining agents follow their strategies $\sigma_{-i}^P$.

Now, let's assume that agent $i$ deviates at $I_{i}$ where $z_{\ell}\in I_{i}$ using the truth-telling strategy corresponding to preference $P_{i}^{\prime }=P^{y}$, \textit{i.e.}, $\sigma _{i}^{P_{i}^{\prime }}(I_i)=\{P^y\}$ for all $I_i\in \mathcal{I}_i$. 
As a result, the set of earliest points of departure for $\sigma _{S_{k+1}}^{P}$ and $\sigma _{i}^{P_{i}^{\prime }}$ includes the information set $I_{i}$ to which $z_\ell$ belongs to. 
Let $P_{N\setminus S_{k+1}}^{\prime }$ be a preference profile for $N\setminus S_{k+1}$ such that $P_{j}^{\prime }=\{P^{y}\}$ for all $j\in N(h_{\ell }^{y})$ and $P_{j}^{\prime }=\{P^{x}\}$ for all $j\in N\setminus (S_{k+1}\cup N(h_{\ell }^{y}))$. 
Define $P^{\prime }=(P_{i}^{\prime },P_{S_{k+1}\setminus\{i\}},P_{N\setminus S_{k+1}}^{\prime })$, and observe that $\{i\in N\mid P^\prime_i=\{P^x\}\}=X_{k+1}^{\ast }\cup (N\setminus (S_{k+1}\cup N(h_{\ell }^{y})))$. 
Thus, $M_{k+1}^{\ast }\subseteq X_{k+1}^{\ast }\cup (N\setminus (S_{k+1}\cup N(h_{\ell }^{y})))$ and $M_{k+1}^{\ast }\in \mathcal{C}^{x}$.
By monotonicity and the definition of $f_{\mathcal{C}^x}$ in (\ref{GMV}), we have
\begin{equation}\label{f(P')=x rebis}
f_{\mathcal{C}^x}(P^\prime)=x.
\end{equation}
Since $\Gamma$ and the type-strategy profile induce $f_{\mathcal{C}^x}$, we can express this as
\begin{equation}\label{f(P')=g rebis}
f_{\mathcal{C}^x}(P^\prime)=g(z^{\Gamma }(z_{0},(\sigma _{i}^{P_{i}^{\prime }},\sigma_{S_{k+1}\setminus \{i\}}^{P},\sigma _{N\setminus S_{k+1}}^{P^{\prime }}))). 
\end{equation}
Combining conditions (\ref{f(P')=x rebis}) and (\ref{f(P')=g rebis}), we find that
\begin{equation*}
x=g(z^{\Gamma }(z_{0},(\sigma_{i}^{P_{i}^{\prime }},\sigma_{S_{k+1}\setminus \{i\}}^{P},\sigma _{N\setminus S_{k+1}}^{P^{\prime }}))).
\end{equation*}
Since, by definition of $P^{\prime }$, $z_{\ell }\in Z_{i}$ belongs to the path induced by the strategy profile $P^{\prime }$, we conclude that  
\begin{equation*}
x=g(z^{\Gamma }(z_{\ell},(\sigma _{i}^{P_{i}^{\prime }},\sigma
_{S_{k+1}\setminus \{i\}}^{P},\sigma _{N\setminus S_{k+1}}^{P^{\prime }}))).
\end{equation*}
Thus, $x$ is the outcome, starting at $z_\ell$, when $i$ chooses $\{P^y\}$ while the remaining agents follow their strategies $\sigma_{-i}^{P^\prime}$. 
Since $y\in o(\sigma^P _{S_{k+1}},\sigma^{P^{\prime }_{i}}_{i})$, $x\in o^{\prime }(\sigma^P _{S_{k+1}},\sigma^{P^{\prime }_{i}}_{i})$, and $x\,P_{i}\,y$, we have that $\sigma _{i}^{P_{i}}$ is not obviously dominant with respect to $\mathcal{S}$ in $\Gamma $ for $i$ with $P_{i}=P^{x}$.
This contradicts the assumption that $(\Gamma,(\sigma _{i}^{P_{i}})_{P_{i}\in \mathcal{P}\,,\,i\in N})$ OSP-implements $f_{\mathcal{C}^{x}}$ with respect to $\mathcal{S}$. 
By proceeding similarly, we can derive a contradiction for the case where $\left\vert X_{1}^{\ast}\right\vert >\left\vert X_{1}^{\prime }\right\vert $. 
Therefore, the necessary condition (i) of Lemma \ref{nuevo} holds for $k+1$ and $(S_1,\dots,S_{k+1})$.

To prove that condition (ii) holds for $k+1$ and $(S_1,\dots,S_{k+1})$ by contradiction, assume there exists $M_{k+1}^{\ast}\in C_{m}^{x}$ such that $X_{k+1}^{\ast }:=M_{k+1}^{\ast }\cap S_{k+1}\notin \mathcal{C}^{x} $, and for all $j\in S_{k+1}\setminus M_{k+1}^{\ast }$, 
\begin{equation*}
X_{k+1}^{\ast }\cup \{j\}\notin \mathcal{C}^{x}.
\end{equation*}
Let $j$ be an agent satisfying this property.
By anonymity relative to $\mathcal{S}$, we can interchange agent $j$ with agent $i$ as identified in the proof of condition (i).
Subsequently, the argument of condition (ii) mirrors that of condition (i), starting from the assumption that agent $i$'s  preference is $P_i=P^x$.
Consequently, the necessary condition (ii) of Lemma \ref{nuevo} holds for $k+1$ and $(S_1,\dots,S_{k+1})$. $\hfill$
$\blacksquare $

\end{document}